\documentclass[journal,comsoc]{IEEEtran}
\usepackage[T1]{fontenc}

\newcommand{\subparagraph}{}
\usepackage{titlesec}
\titleformat*{\subsubsection}{\fontsize{9.5}{10}\bfseries}
\titleformat*{\subsection}{\fontsize{10}{11}\itshape}
\titleformat*{\paragraph}{\fontsize{9}{9.5}\bfseries}

\usepackage[hyphens]{url}

\usepackage{mathtools}

\usepackage{ragged2e}
\usepackage{blindtext, graphicx}
\graphicspath{{Pictures/}{Profile/}} 
\usepackage{enumitem} 
\usepackage{hyphenat} 
\usepackage[tracking=true]{microtype} 
\SetTracking[no ligatures = f]{encoding = *,shape = sc}{120}
\usepackage{amsmath, amsthm, amssymb}
\usepackage{dblfloatfix}
\usepackage{soul}
\usepackage{xcolor}

\soulregister\cite7
\soulregister\ref7
\soulregister\pageref7

\usepackage{cite}

\ifCLASSINFOpdf
\else
\fi
\usepackage{amsmath}
\usepackage[cmintegrals]{newtxmath}
\ifCLASSOPTIONcompsoc
  \usepackage[caption=false,font=footnotesize,labelfont=sf,textfont=sf]{subfig}
\else
  \usepackage[caption=false,font=footnotesize]{subfig}
\fi

\begin{document}
%
\title{A System-Level Framework for Analytical and Empirical Reliability Exploration of STT-MRAM Caches}
%
%
%

\author{Elham~Cheshmikhani,~Hamed~Farbeh,\IEEEmembership{}~and~Hossein~Asadi,~\IEEEmembership{Senior Member,~IEEE}
\thanks{E. Cheshmikhani and H. Asadi (corresponding author) are with the Department of Computer Engineering, Sharif University of Technology, Tehran 1115511365, Iran.\protect\\
			E-mail: elham.cheshmikhani@sharif.edu; asadi@sharif.edu}
\thanks{H. Farbeh is with the Department of Computer Engineering, Amirkabir University of Technology, Tehran  1591634311, Iran.\protect\\
			E-mail: farbeh@aut.ac.ir.}
\thanks{Manuscript received September 23, 2018, revised March 8, 2019, accepted May 8, 2019.}}

%
%

\markboth{IEEE Transactions on Reliability,~Vol.~xx, No.~xx, JUNE~2019}%
{Shell \MakeLowercase{\textit{et al.}}: Bare Demo of IEEEtran.cls for IEEE Communications Society Journals}
%



\maketitle

\begin{abstract}
$Spin$-$Transfer$~$Torque$~$Magnetic$~$RAM$ (STT-MRAM) is known as the most promising replacement for SRAM technology in 
			large $Last$-$Level~Cache$ memories (LLCs).
			Despite its high-density, non-volatility, near-zero leakage power, and immunity to radiation as the major advantages, STT-MRAM-based cache memory suffers from high error rates mainly due to {$retention~failure$}, {$read~disturbance$}, and {$write~failure$}.
Existing studies are limited to estimate the rate of $only$ one or two of these error types for STT-MRAM cache. 
However, the overall vulnerability of STT-MRAM caches, which its estimation is a must to design cost-efficient reliable caches, has not been offered in none of previous studies.

In this paper, we propose a system-level framework for reliability exploration and characterization of errors behavior in STT-MRAM caches.
To this end, we formulate the cache vulnerability considering the inter-correlation of the error types including retention failure, read disturbance, and write failure as well as the dependency of error rates to workloads behavior and $Process~Variations$ (PVs). 
Our analysis reveals that STT-MRAM cache vulnerability is highly workload-dependent and varies by orders of magnitude in different cache access patterns.
Our analytical study also shows that this vulnerability divergence significantly increases by process variations in STT-MRAM cells.
To take the effects of system workloads and PVs into account, we implement the error types in gem5 full-system simulator.
The experimental results using a comprehensive set of multi-programmed workloads from SPEC CPU2006 benchmark suite on a quad-core processor show that the total error rate in a shared 
STT-MRAM 
LLC varies by 32.0x for different workloads.
A further 6.5x vulnerability variation is observed when considering PVs in the STT-MRAM cells.
In addition, the contribution of each error type in total 
LLC vulnerability highly varies in different cache access patterns and moreover, error rates are differently affected by PVs.
The proposed analytical and empirical studies can significantly help system architects for efficient utilization of error mitigation techniques and designing highly reliable and low-cost STT-MRAM 
LLCs.

\end{abstract}


%
\IEEEpeerreviewmaketitle

\section{Introduction}
%
%
%
%
\IEEEPARstart{T}{echnology-scaling} trend in recent years has led to emerging challenges for the traditional SRAM-based 
			large $Last$-$Level~Caches$ (LLCs)
			including unreliability, leakage power, and low density 
			in multi- and many-core processors
			~\cite{
			itrs,ogden2017impact, farbeh2016floating, ghaemi2019sleepy}.
To tackle these challenges, extensive ongoing industrial and academic research efforts have focused on replacing SRAMs with emerging \textit{$Non$}-$Volatile$~$Memories$ (NVMs)~\cite{wu2016temperature,Chen2016,ghaemi2019sleepy,vatajelu2017challenges, salkhordeh2016operating}.
According to the recent industrial reports, \textit{$Spin$}-\textit{$Transfer~Torque$} \textit{$Magnetic~Random$} \textit{$Access~Memories$} (STT-MRAMs) are the most promising technologies to substitute SRAMs in 
LLCs
~\cite{wu2016temperature, wang2016memres,Chintaluri2015, imani2016approximate, kim2016exploration}.

			STT-MRAM caches benefit from non-volatility, higher-density, near-zero leakage power, and immunity to radiation-induced particle strikes~\cite{zuo2018write,salkhordeh2019analytical,mittal2015survey, farbeh2018cache}.
			However, STT-MRAM reliability is a major challenge for its applicability in 
			LLCs.
\textit{$Retention~failure$} (i.e., the content of a cell flips during its idle time)~\cite{Chen2016, kim2016exploration, Chintaluri2015}, \textit{$read~disturbance$} (i.e., unintentional flip of a cell due to applying read current during a read access)~\cite{fong2014failure,aliagha2019react,kang2015yield, wang2015selective}, and \textit{$write~failure$} (i.e., unsuccessful write operation due to inability of a cell to switch) ~\cite{farbeh2016floating,eli-aspdac,choi2017nvm} are the main sources of errors in STT-MRAM 
LLCs.
Stochastic switching behavior of magnetic field direction in STT-MRAM cells is the source of the mentioned error types~\cite{zhao2012spin, farbeh2016floating, lakys2012self}.
Besides the device-level parameters and characteristics of STT-MRAM cells, the rates of these errors are affected by two major sources: a) system-level parameters, e.g., workloads, which affect the memory content and access patterns \cite{salkhordeh2018reca,bor2017tdsc,eli-aspdac, kishani2018modeling} and
b) physical parameters, which are $Process~Variations$ (PVs) that differently (sometimes oppositely) affect these error rates~\cite{imani2016approximate, kim2016exploration, kang2015reconfigurable, 12-EDCC-sun2012process}.

			From the workload perspective, the diversity in cache access patterns and the content of data blocks in different workloads play an important role in the rate of the retention failure, read disturbance, and write failure.
Retention failure probability in cache blocks with larger gap between consecutive accesses is higher, whereas more frequently-accessed blocks experience lower failure probability.
However, the occurrence probability of read disturbance and write failure per unit of time is higher for more frequently-accessed blocks.
On the other hand, the rate of read disturbance and write failure directly depends on the content of data blocks.
Read disturbance error is unidirectional and only $unintentional$ ${1\rightarrow 0}$ transitions are probable. 
Write failure, on the other hand, is only probable when the content of the original and updated values are different in the cache.
			Therefore, read disturbance and write failure rates are affected by the number of `1's in a block and the hamming distance between the content of a cache block and the updated data block, respectively.

			In addition to the system workloads, PVs can significantly change the error rates by deviating the physical parameters of a cell from their nominal values. 
			A deviation changes the rate of the mentioned error types differently and even in opposite directions~\cite{12-EDCC-sun2012process,kang2015reconfigurable, zhao2012spin}.
Considering an STT-MRAM cache as an array of STT-MRAM cells, the overall retention failure, read disturbance, and write failure rates depend on the interaction of process variations effects among the STT-MRAM cells.


Although the error rate of a single STT-MRAM cell can be estimated using its physical parameters,
the vulnerability estimation of the entire STT-MRAM cache requires system-level analysis considering the workloads behaviors as well as PV effects.
A recent study has reported only the rate of write failure and read disturbance in STT-MRAM cache and shown that these rates are strongly workload-dependent~\cite{eli-aspdac, EDCC}.
To the best of our knowledge, none of the previous studies neither have addressed the workloads- and PVs-dependent vulnerability of STT-MRAM caches considering all types of errors, nor reported the cache reliability considering the conflicting sources of errors.
Therefore, to achieve an acceptable level of reliability in STT-MRAM cache with affordable costs, it is necessary to \textbf{a)} estimate the vulnerability of the cache to each error type as well as the total cache vulnerability considering the conflicting dependencies of error types to different factors and \textbf{b)} utilize effective mitigation techniques for each error type according to its contribution in the total cache vulnerability.

In this paper, we propose a system-level framework for early exploration of STT-MRAM cache reliability, which considers \textbf{a)} workload-dependent cache access patterns and data content, \textbf{b)} PV affected STT-MRAM cell vulnerability to errors, and \textbf{c)} inter-correlation between the rate of three-mentioned errors, i.e., retention failure, read disturbance, and write failure, for reliability exploration.
To this end, we first formulate the vulnerability of STT-MRAM caches to each error type by extrapolating the error probability of PV affected cache blocks from nominal error probability of a single STT-MRAM cell.
Then, we deduce the error probability of a PV affected STT-MRAM cache from the blocks error probability. 
Our formulation is then extended to integrate probabilities of the three differently-originated and exclusively-occurred error types into a single cache vulnerability equation.

The inputs of our architecture-level formulas are twofold: 1) device-level parameters of single STT-MRAM cell, e.g., thermal stability factor, critical switching current, tunneling spin polarization, and write/read current, which are derived from several industrial and technical reports, 2) system-level parameters derived from workload behavior, e.g., frequency and sequence of read and write operations, data content, and cache block idle intervals.

We present a framework integrated with gem5 full-system simulator~\cite{gem5} to extract and analyze the reliability characteristics of STT-MRAM LLCs based on our formulations.
This framework provides a high flexibility in terms of STT-MRAM cells physical and circuit-level parameters as well as cache configurations.
For a careful evaluation of the STT-MRAM cache vulnerability, the rate calculation for each error type is implemented in the gem5 based on a) the number of accesses, b) idle times, c) data content, and d) cell transitions.
The evaluations have been conducted for an STT-MRAM 
LLC
shared across cores of a quad-core processor running a set of multi-programmed workloads from SPEC CPU2006 benchmark suite~\cite{spec2006}. 
The results show that the rate of each error type as well as the total error rate significantly varies for different workloads.
The workload dependency of the errors resulted in a minimum of 6.8x variation in the rate of three types of errors.
This value is 32.0x for total cache failure probability, on average.
Our observations also reveal that PVs increase the occurrence probability of all error types by more than 8x for all workloads.

			
			The \textbf{main}~\textbf{contributions} of this paper are as follows:
			
			\begin{enumerate}
 				 \item This is the first study that formulates the vulnerability of STT-MRAM caches to retention failure, read disturbance, and write failure errors based on the workload-dependent cache behavior and by extrapolating the nominal error rate of a single STT-MRAM cell.
			 	 \item We formulate the total reliability of STT-MRAM caches by proposing an approach to integrate the cache unreliability per unit of time (for retention failure), per read access (for read disturbance), and per write access (for write failure) into a unified cache failure probability.
			 	 \item We investigate the effects of PVs on the rate of all three error types as well as the total error rate in STT-MRAM cache. To the best of our knowledge, this is the first study that investigates the system-level reliability impacts of STT-MRAM physical parameters affected by PVs.
			 	 \item We present a framework integrated with gem5 full-system simulator~\cite{gem5} to extract and analyze the reliability characteristics of STT-MRAM 
			 	 LLCs
			 	 based on our formulations. This framework provides a high flexibility in terms of STT-MRAM cells physical and circuit-level parameters as well as cache configurations.
			 	 The proposed framework supports both perpendicular and in-plane STT-MRAM technologies. 
			 	\item We investigate the dependency between workloads behavior and the rate of errors in a STT-MRAM 
				LLC
				shared across cores of a quad-core processor.
			 	Our study reveals that retention failure, read disturbance, and write failure vary by 25.4x, 6.8x, and 15.1x in different workloads, respectively.
			 	A 32.0x variation is also observed for the total error rate including all types of errors. 
			 	These observations indicate that the cache access pattern and data content not only extremely differentiate the rate of errors, but also their effect on the rate of three error types is significantly different.
			 	\item We investigate the effects of PVs on the 
			 	LLC
			 	error rates and show that PVs increase the rate of retention failure, read disturbance, and write failure by 32.5x, 9.0x, and 16.7x, respectively.
			 	In addition, the total error rate is increased by 6.5x.
			 	These observations reveal that the susceptibility of the error types to PVs is highly different and the effect of PVs on the cache reliability is mostly determined by an error type with the highest contribution in the total error rate.
			\end{enumerate}

			The rest of this paper is organized as follows. Section II describes the basics of STT-MRAM technology and its reliability challenges. In Section III, the observations and motivations for this work are discussed. The details of the proposed formulations and framework are presented in Section IV. Section V gives the simulation setup and evaluation results. A discussion on the existing reliability improvement techniques and suggested guidelines based on our study is presented in Section VI. Finally, we conclude the paper in Section VII.

\section{Preliminaries}
			\label{sec:PRELIMINARIES}
			In recent years, beside STT$-$MRAM,~\textit{Spin}-\textit{Orbit}~\textit{Torque}~\textit{Magnetic}~\textit{RAM} (SOT$-$MRAM) is introduced as an alternate generation of MRAM technology.
SOT$-$MRAM is based on the three-terminal MTJ and uses \textit{Spin} \textit{Hall} \textit{Effect} to switch.
It tries to overcome high write latency and energy of STT$-$MRAM by separating the read path from the write path.
This separation also avoids read disturb in the cell. Despite of these advantages, unlike STT$-$MRAM that requires only one access transistor for both read and write, SOT$-$MRAM needs two transistors for read and write operations.
This makes the STT$-$MRAM denser than SOT$-$MRAM.
On the other hand, STT$-$MRAM memory is more mature and is designed and manufactured by big companies such as Toshiba~\cite{Toshiba}, TSMC~\cite{tsmc}, and Samsung~\cite{samsung} as a prototype and is also used in recent commercial products. Although SOT$-$MRAM has some advantages over STT$-$MRAM and investigating its reliability characteristics can be an interesting research track, it takes a long way for SOT$-$MRAM to be commercialized and this study focuses only on STT$-$MRAM technology. In this section, we explain the structure of STT-MRAM cells and the mechanisms of read and write operations in STT-MRAM cells. Then, we focus on the sources of errors in STT-MRAM cells and discuss the dependency of the rate of these errors to STT-MRAM cells configuration.

				\subsection{STT-MRAM Cell Basics}
\begin{figure}[t]
				\centering
				\subfloat[]{\includegraphics[width=0.35\linewidth]{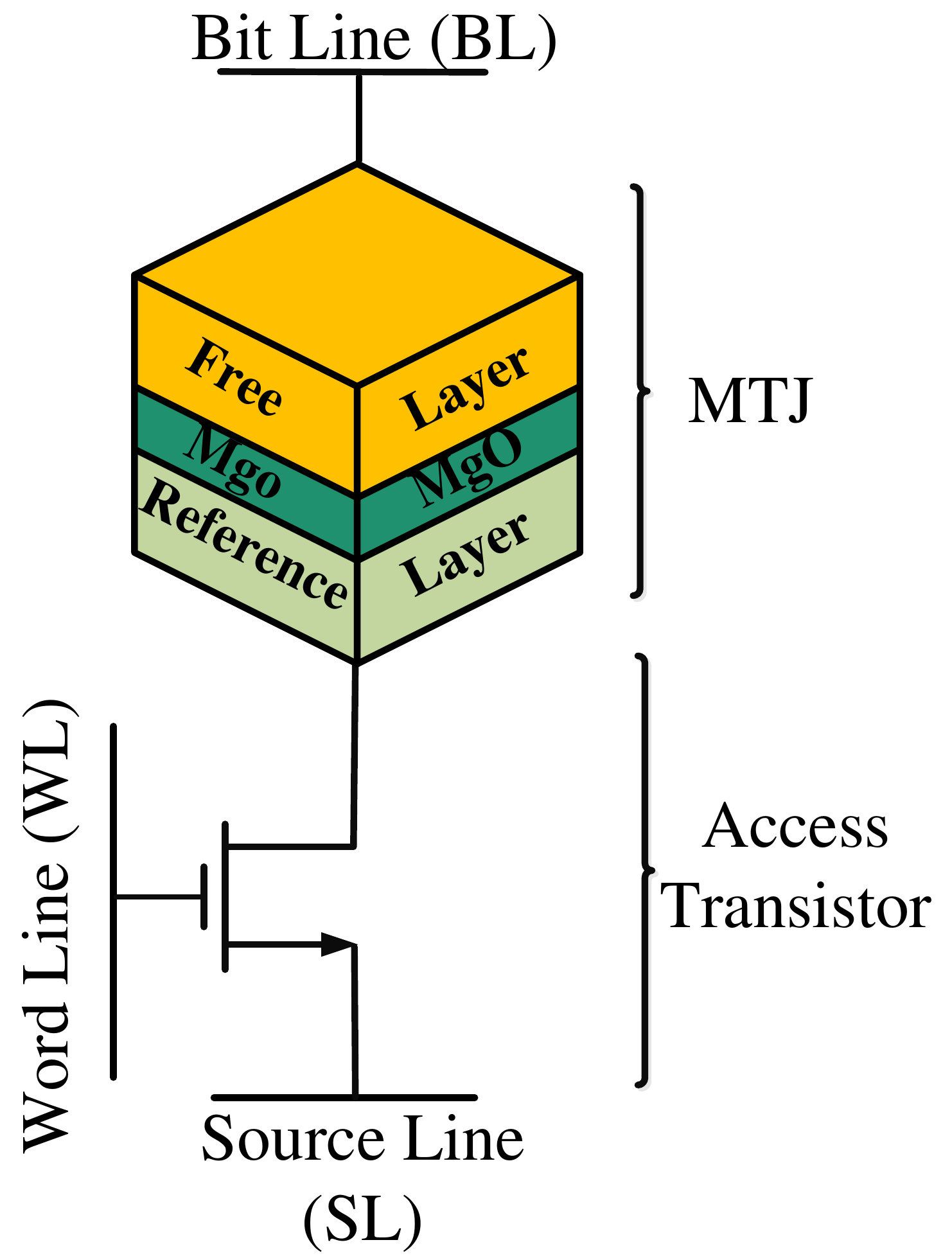}}
				\hspace{5pt}
				\subfloat[]{\includegraphics[width=0.59\linewidth]{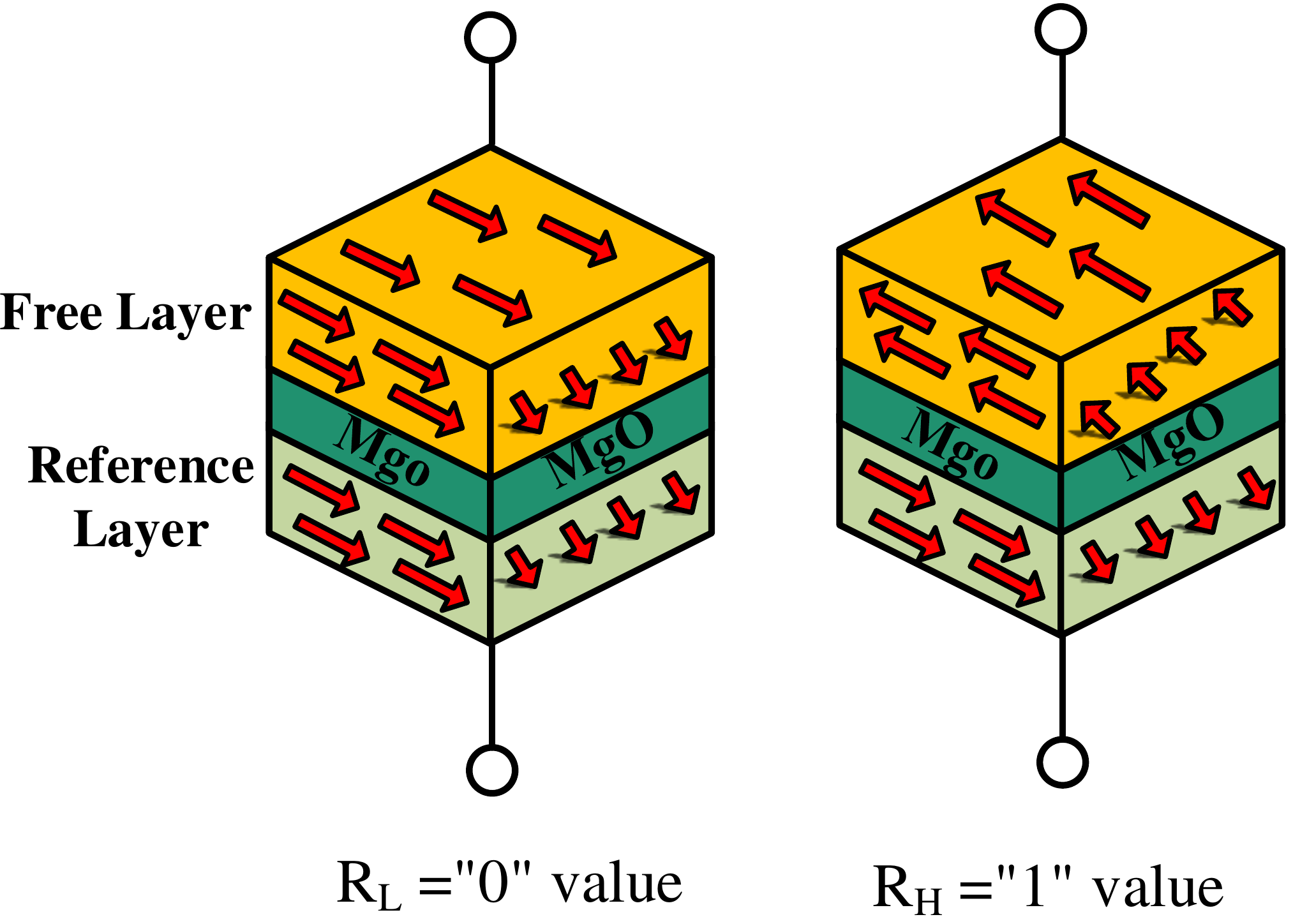}}
				\caption{STT-MRAM cell schematic: (a) 1T1J STT-MRAM cell structure and (b) MTJ low and high resistance states.}\vspace{-10pt}
				\label{fig:1}
			\end{figure}

			An STT-MRAM cell comprises a storage component, named \textit{$Magnetic~Tunnel~Junction$} (MTJ), to store data values, and an NMOS access transistor.
The structure of this cell, known as 1T1J STT-MRAM, is shown in Fig.~\ref{fig:1}(a). 
An MTJ consists of two ferromagnetic layers separated by a thin oxide barrier layer~\cite{wu2016temperature, farbeh2018cache, Eli-TC}. 
These layers as depicted in Fig.~\ref{fig:1}(b) are \textit{$reference$}, \textit{$free$}, and $tunnel~barrier$.
The barrier layer made of crystallized $Magnesium~oxide$ (MgO) insulates the electrons movement from the reference layer to the free layer and vice versa.
The reference layer has a fixed magnetic field direction, while the magnetic field direction of the free layer can be changed by applying a magnetic force \cite{khvalkovskiy2013basic,14-EDCC-apalkov2013spin, kang2015yield}.
			
			As the STT-MRAM technology is based on magnetic charge instead of electrical charge, the magnetic field direction of the MTJ ferromagnetic layers determines the cell content. The relative direction of magnetic field in the free and reference layers (anti-parallel or parallel directions) causes two different resistances, i.e., $R_{High}$ and $R_{Low}$ ($R_{H}$ and $R_{L}$), in MTJ  as shown in Fig.~\ref{fig:1}(b).
$R_{H}$ and $R_{L}$ represent high and low MTJ resistances, respectively, and are distinguished based on the voltage value between the $bit~line$ (BL) and $source~line$ (SL) terminals on a read operation. $Tunneling~Magneto~Resistance$ (TMR) ratio parameter, which is defined as $TMR$ = ($R_{H}$ - $R_{L}$)/$R_{L}$, shows the ratio between these resistances and indicates the cell state~\cite{15-EDCC-zhao2011design, zhao2012spin
}. 
When the MTJ is on parallel (or anti-parallel) state, the MTJ resistance is low (or high), which represents that `0' (or `1') logic is stored in the cell. The MTJ states and its logic values are depicted in Fig.~\ref{fig:1}(b).
		
			\subsection{Read and Write Operations}
\begin{figure}[t]
				\centering
				\subfloat[]{\includegraphics[width=0.47\linewidth]{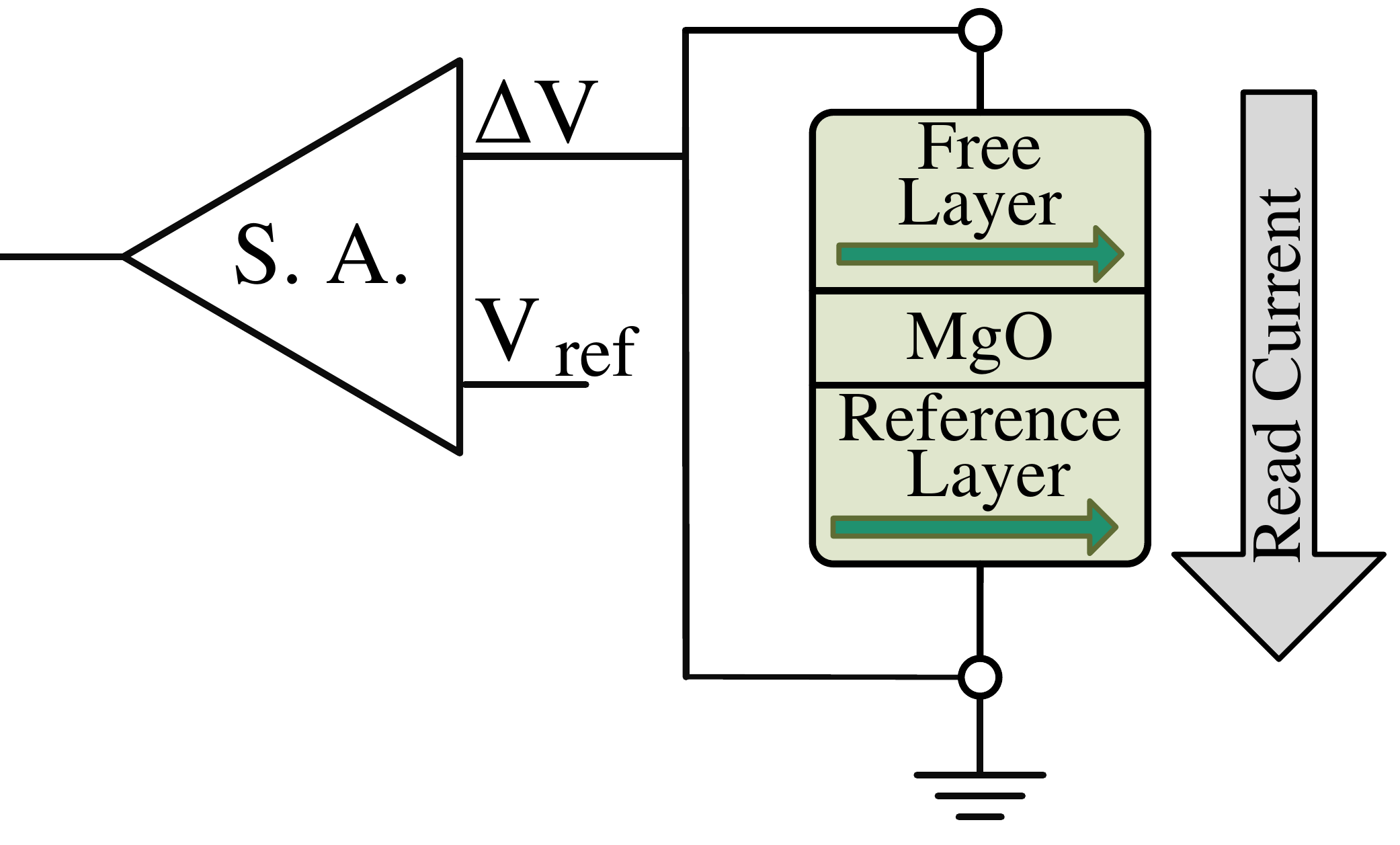}}
				\vspace{10pt}
				\hspace{10pt}
				\subfloat[]{\includegraphics[width=0.45\linewidth]{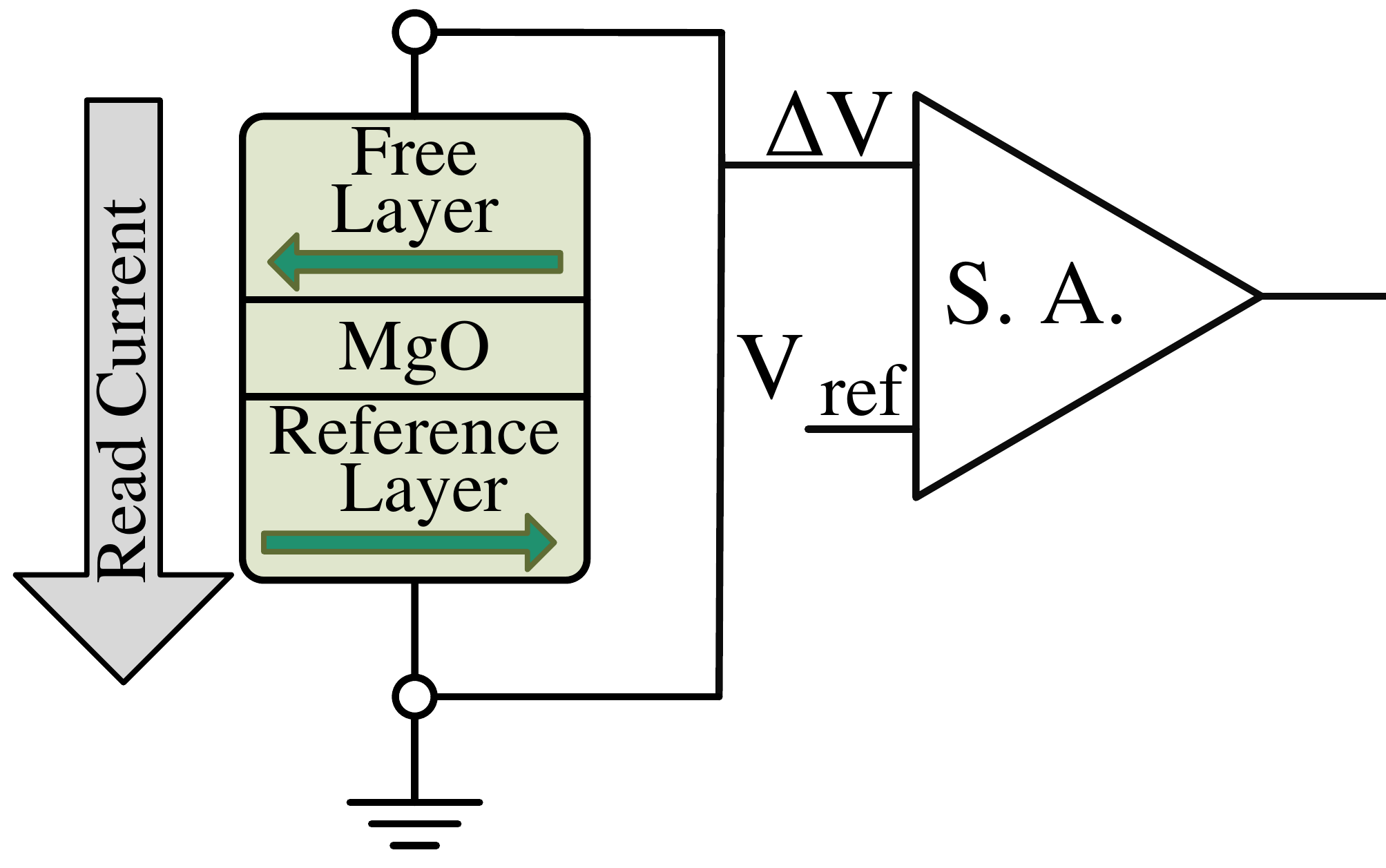}}\vspace{-3pt}
				\hspace{10pt}
				\subfloat[]{\includegraphics[width=0.45\linewidth]{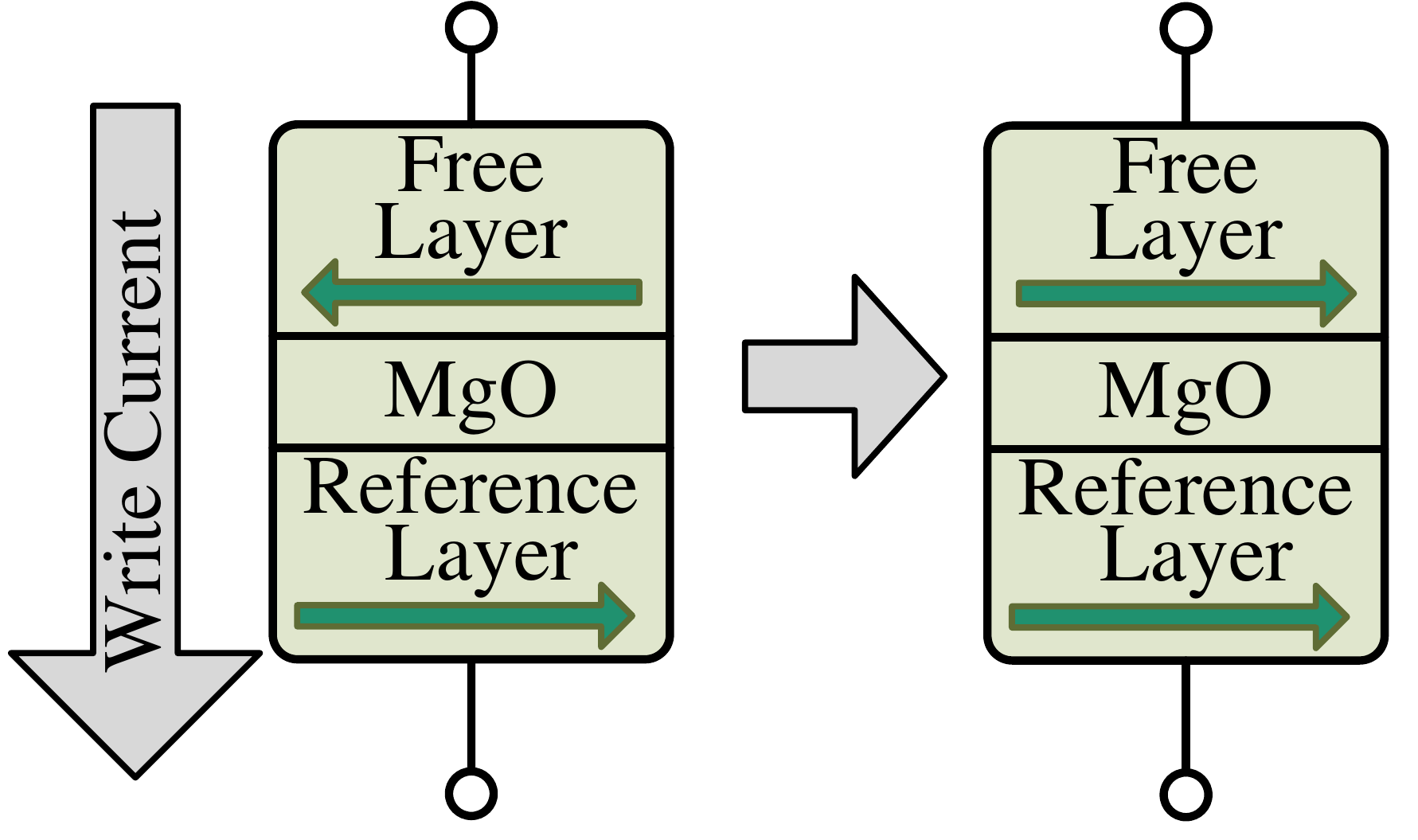}}
				\hspace{10pt}
				\subfloat[]{\includegraphics[width=0.45\linewidth]{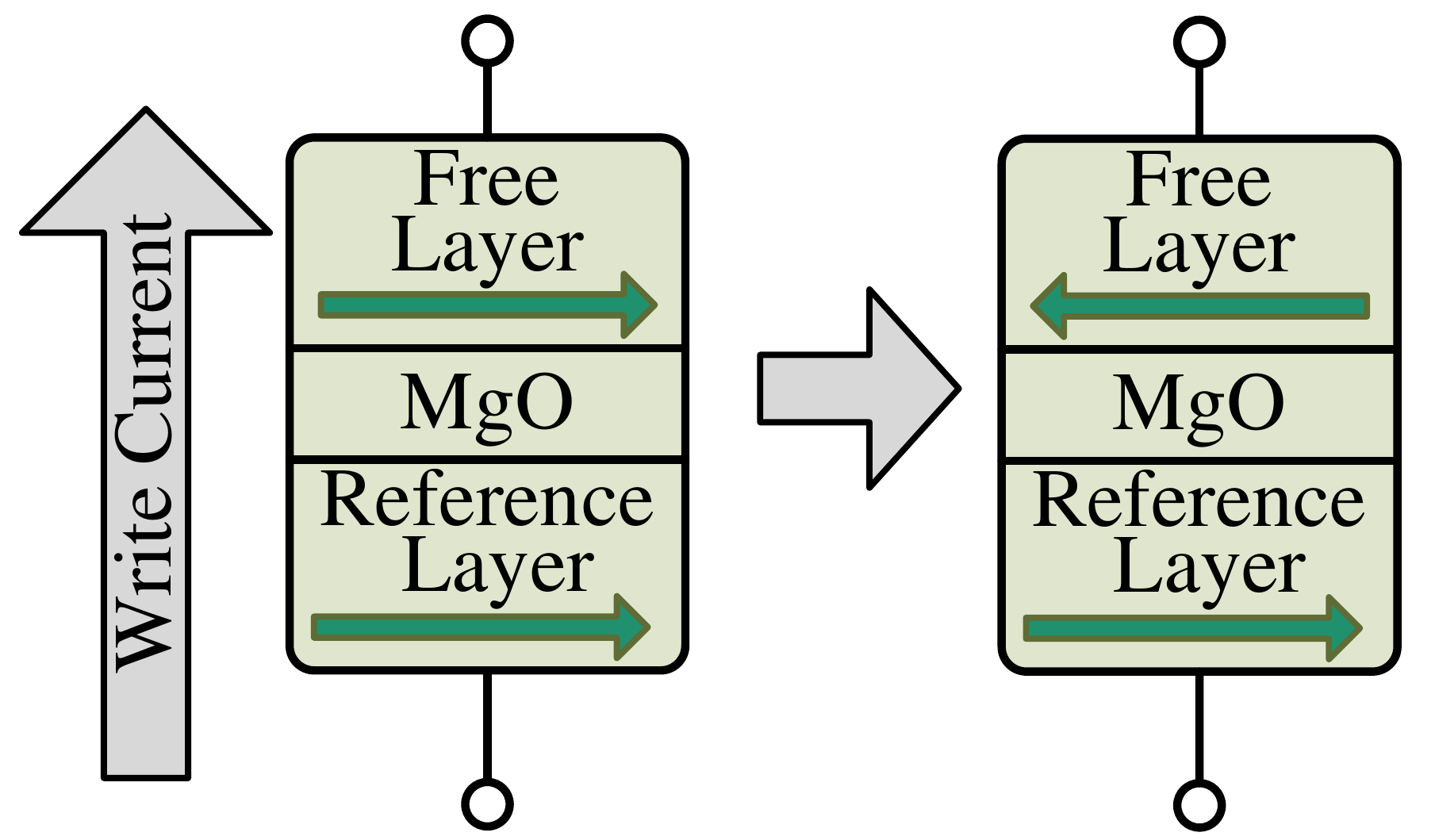}}\vspace{-3pt}
				\caption{STT-MRAM read and write operations: (a) reading `0', (b) reading `1', (c) writing `0', and (d) writing `1'.}\vspace{-10pt}
				\label{fig:basics}
			\end{figure}

			As mentioned, the resistance of MTJ ($R_H$ or $R_L$) shows the value in an STT-MRAM cell. To read the stored value in a cell, first the $word~line$ (WL) is set to turn on the access transistor. Then, a small read current ($I_{read}$) should be applied to the STT-MRAM cell~\cite{kim2016exploration, itrs}.
			In (\ref{eq:1}) and (\ref{eq:2}), the {$bit~line~voltage$} ($V_{BL}$) in a 1T1J STT-MRAM cell is shown.\vspace{-15pt}
			
				\begin{flalign}
			\label{eq:1}
			\resizebox{.55\linewidth}{!}{$ V_{BL-Low} = {I_{read}\times(R_L+R_{NMOS})}$}    \phantom{\hspace{1.2cm}}
				\end{flalign}
			\begin{flalign}
			\label{eq:2}
			\resizebox{.55\linewidth}{!}{$ V_{BL-High} = {I_{read}\times(R_H+R_{NMOS})}$}    \phantom{\hspace{1.2cm}}
			\end{flalign}	
			Where, $V_{BL-Low}$ and $V_{BL-High}$ are the bit line voltages when the MTJ is at low and high resistance states, respectively.
			In these equations, $R_L$ and $R_H$ are the low and high MTJ resistance, respectively, $R_{NMOS}$ is the resistance of NMOS access transistor, and $I_{read}$ is read current~\cite{14-zazad-eken2014novel, eli-date}. 
			By applying $I_{read}$ to an STT-MRAM cell, a voltage is generated between the bit line and source line. 
			By comparing the $V_{BL}$ with a {$reference~voltage$} ($V_{REF}$), the MTJ resistance state can be read out~\cite{ mittal2017survey, 15-EDCC-zhao2011design}.		
			As shown in Fig. \ref{fig:basics}(a) (Fig. \ref{fig:basics}(b)), if the sensed value is higher (lower) than the reference voltage, it means that the resistance of MTJ is low (high) and the cell contains `0' (`1') value.  
			
			Write operation is more complicated than read operation due to requiring to change the MTJ resistance. 
MTJ resistance changes if the magnetic field direction of the free layer flips. 
To this end, a write current is applied to the source line or bit line to write `0' or `1', as shown in Fig. \ref{fig:basics}(c) and Fig. \ref{fig:basics}(d), respectively.
Consequently, electrons spin in the free layer orient in the same direction or opposite direction of the reference layer magnetic field, based on the direction of the applied current.
This phenomenon causes a spin-polarized current. 
When the amount of spin-polarized current exceeds a threshold value, the magnetic field direction of the free layer switches. 
This is the time when the MTJ content flips and a value is written into the cell.
Switching the magnetic field direction from anti-parallel to parallel (or from parallel to anti-parallel) leads electrons to flow from the reference layer to the free layer (or vice versa)~\cite{
mittal2017survey, 15-EDCC-zhao2011design,kang2014variation, farbeh2018cache}. 

\subsection{STT-MRAM Reliability} 
			Reliability of STT-MRAM cells is threatened by retention failure, read disturbance, and write failure. $Retention~failure$ occurs when a cell is idle (a cell that is not being read nor written) and its content flips stochastically. $Read~disturbance$ occurs during a read operation when the content of a cell changes unintentionally. A $write~failure$ occurs when the content of a cell is not switched by the current applied during the write operation.
The origin of all these errors is the stochastic switching behavior of STT-MRAM cells~\cite{Pajouhi2016JETC,Na-TCAS-II-16,Ran2016JSA, eli-date}.

			Read disturbance and  write failure are the main reliability threats in 22nm technology node~\cite{naeimi2013intel}.
While the process technology scales down, it is predicted that the retention failure probability increases exponentially, read disturbance probability increases in lower rate than retention failure probability, whereas the rate of write failure decreases.
Therefore, the total error rate considering all three sources of errors is significantly increased by technology downscaling, in which the retention failure is the main contributor.

			Because of a retention failure, the content of an idle STT-MRAM cell flips accidentally without passing any current through it.
			The retention failure probability in time interval \textit{t} is calculated according to (\ref{eq:3})~\cite{15-EDCC-zhao2011design, naeimi2013intel}.
				\begin{flalign}
			\label{eq:3}
			\resizebox{.8\linewidth}{!}{$ P_{Retention-Failure} = 1- exp({-t }\times {exp(-\Delta)})$}
			\end{flalign}
			Where, $t$ is the cell idle time and $\Delta$ is $thermal~stability~factor$ of a STT-MRAM cell.
			The retention failure probability increases in larger values of $t$ and/or smaller values of $\Delta$. 
			The thermal stability factor of a cell is calculated according to (\ref{eq:4})~\cite{naeimi2013intel}.\vspace{-10pt}
		
			\begin{flalign}
			\label{eq:4}
			\resizebox{.18\linewidth}{!}{$\Delta = \frac{E_{b}}{ K T}$}
			\end{flalign}
			Where, \textit{E$_b$} is barrier energy, \textit{K} is Boltzmann constant, and \textit{T} is temperature in Kelvin.

			During a read operation in a STT-MRAM cell, content of the cell can change unintentionally due to the read current passing through the cell, causing a read disturbance error. The direction of the read current, which flows through the MTJ layers in a read operation, is the same as the direction of the current for writing either `1' or `0' in a write operation. 
			The current in read operation is much lower than the current in the write operation~\cite{farbeh2016floating}.
			However, this low current can cause an unintentional flip in the cell during a read operation and erroneously change its content.
			The switching probability of an STT-MRAM cell during a read operation is according to (\ref{eq:5})~\cite{lakys2012self, eli-date, naeimi2013intel}.
			\begin{equation}
			\begin{multlined}
			\label{eq:5}
			\shoveright[2cm]{P_{Read-Disturbance}=1- exp(\frac{-t_{read}}{\tau}\times}\\
			\shoveleft[2.9cm]{exp(\frac{\Delta(I_{read}-I_{C_0})}{I_{C_0}}))} 
			\end{multlined}
			\end{equation}
			Where, \textit{$\tau$} is attempt period equal to 1ns, \textit{I$_{read}$} is read current, \textit{$I_{C_{0}}$} is critical switching current, \textit{t$_{read}$} is read pulse duration, and $\Delta$ is thermal stability factor.

			Write failure as another reliability challenge occurs when the magnetic field direction of the free layer in MTJ cannot be switched during the interval where write pulse is applied \cite{Pajouhi2016JETC, eli-date}. 
			The MTJ switching time depends on various parameters, e.g., MTJ switching current, process variations, thermal fluctuations, and switching pulse width. The occurrence probability of a write failure for a STT-MRAM cell is according to (\ref{eq:6})~\cite{wu2016temperature, 15-EDCC-zhao2011design, 14-zazad-eken2014novel}.
			\begin{equation}
			\begin{multlined}
			\label{eq:6}
			 P_{Write-Failure} = exp( -t_{write}\times \\
			\shoveright[1cm]{\frac{2 \times \mu_{\beta}\times p\times(I_{write}-I_{C_0})}{c+\log_{e}(\pi^2\times\Delta/4)\times (e\times m\times (1+p^2))})} 
			\end{multlined}
			\end{equation}		
			Where, \textit{I$_{write}$} is write current, \textit{c} is Euler constant, \textit{e} is electron charge, \textit{m} is magnetic momentum of the free layer, \textit{p} is tunneling spin polarization, \textit{$\mu$$_{\beta}$} is Bohr magneton, \textit{t$_{write}$} is write pulse duration, and $\Delta$ is thermal stability factor.


Beside write failure, retention failure, and read disturbance errors, there are three other factors that may affect the STT-MRAM error rate as follows: \textbf{a)}~\textbf{false}~\textbf{read} in which the sense amplifier circuit is unable to correctly determine the content of STT-MRAM cell. This error is because of peripheral CMOS circuitry and it is not specific to STT-MRAM technology; \textbf{b)}~\textbf{limited} \textbf{endurance}, which is due to oxide barrier breakdown of STT-MRAM cells. Although some studies addressed this challenge, several recent industrial reports have illustrated an unlimited endurance for STT-MRAM cells compared to the ten-year lifetime of digital system; and \textbf{c)}~\textbf{radiation-induced}~\textbf{transient}~\textbf{errors} in peripheral CMOS circuitry, which is not related to STT-MRAM technology. While retention failure, read disturbance, and write failure are inter-correlated in STT-MRAM cell, limited endurance is no longer a reliability concern and false read and transient error have never been known as a source of STT-MRAM vulnerability.

			\section{Motivation}
			The reliability challenge of STT-MRAM cells due to retention failure, read disturbance, and write failure becomes more threatening when considering process variations.
			On the other hand, workloads and data patterns affect the rates of these errors. 
			In the following, we first discuss the major reliability challenges by exploring the workloads effects on the error rates. Then, we explore the effects of process variations on these workload-dependent errors.

			\subsection{Effects of Workloads and Data Patterns}	
			Retention failure rate depends on the data access patterns of the workload in a system.
			For example, retention failure probability for a cache block not accessed for a long time is much higher than a cache block frequently accessed in a short interval.
			On the other hand, a retention failure in a cache block is masked if the block is overwritten before it is read.
			Hence, the errors occurred in intervals between a read/write access and its subsequent write access has no effect in data integrity and these intervals are beyond of vulnerable times.
			As a result, the retention failure rate varies for different workloads, different phases of a single workload, and even different cache blocks in a given time slot.

			As shown in Fig. \ref{fig:basics}(a), the read current applied to STT-MRAM cells for read operation is always in a predetermined direction (from SL to BL or vice versa). In other words, read current is predeterminedly in the direction of writing either `0' or `1'. 
			Therefore, only STT-MRAM cells containing `1' (or `0') are vulnerable to read disturbance and the other cells containing `0' (or `1') are not affected.
			This $unidirectional$ read current makes the occurrence probability of read disturbance in a cache block to be content-dependent.

			On the other hand, the total read disturbance rate in a cache during a workload execution depends on the number of read accesses to the cache.
			Obviously, both the content of cache blocks (the number of `0's and `1's) and the number of read accesses can have a large variation for different workloads.
			Therefore, besides its dependency to physical characteristics of STT-MRAM cells and the circuit-level parameters of the cache, the read disturbance rate is highly workload-dependent.

			Same as the rate of retention failure and read disturbance, write failure rate is highly affected by workloads behavior. 
			On a write operation in a cache block, a subset of STT-MRAM cells that their content need to switch are vulnerable to write failure, since it is probable that these cells may not switch during applying the write current.
			No write failure is probable in other cells, since their content is the same as the updated value.
			In this regard, the write failure rate depends on not only the number of write requests, but also the similarity between the content of cache blocks and the value of updated data.

\begin{figure}[t]
				\centering
				\includegraphics[width=1\linewidth]{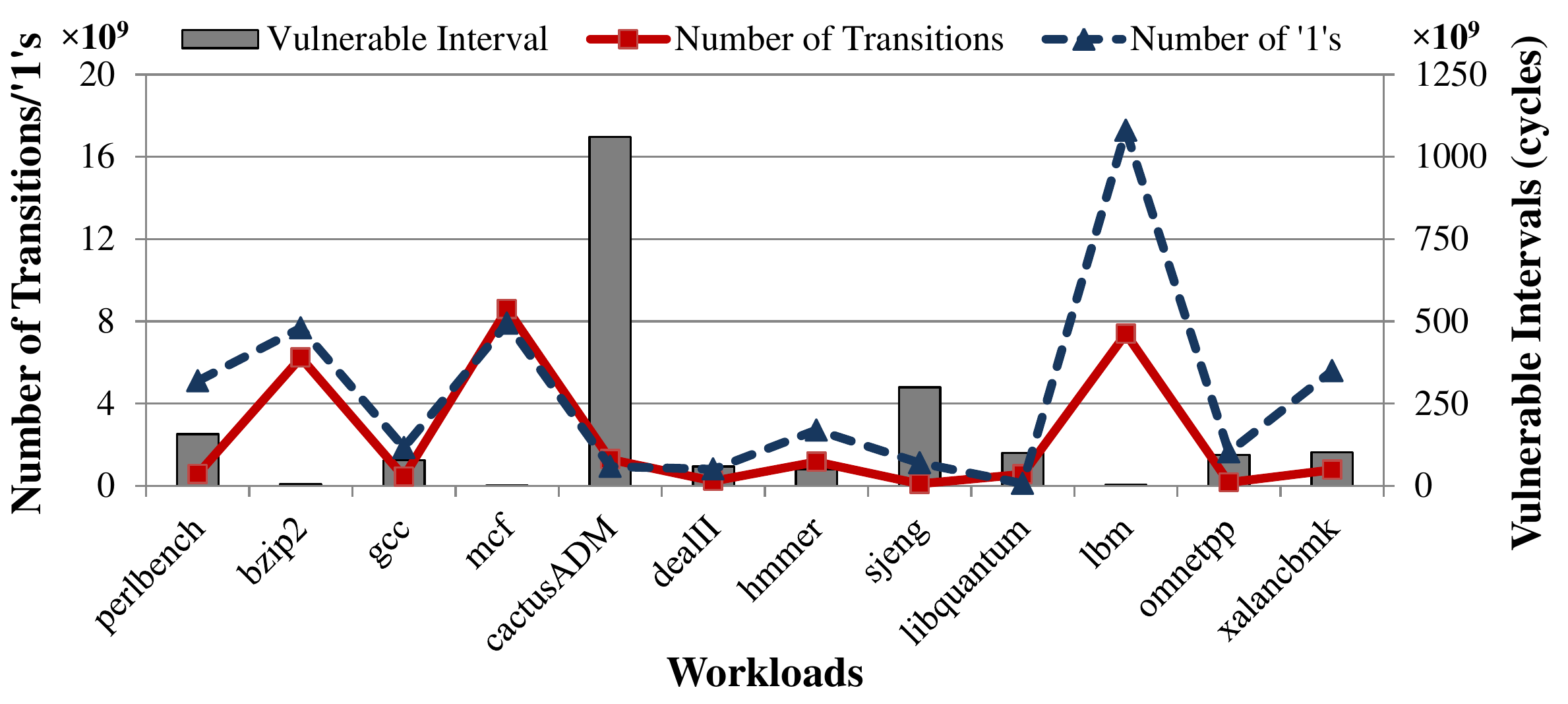}\vspace{-10pt}
				\caption{Workload dependency of parameters affecting retention failure, read disturbance, and write failure rate.}\vspace{-10pt}
				\label{fig:motive}
			\end{figure}

			In addition, for cells that need to switch on a write operation, the probability of unsuccessful switching in ${0\rightarrow 1}$ transitions is by more than two orders of magnitude higher than in ${1\rightarrow 0}$ transitions\cite{ZAZADTPDS, Na-TCAS-II-16}.
			This $asymmetric$ write failure rate for the directions of cell switching makes the unreliability of STT-MRAM caches even more content-dependent.		
			Based on the above discussions, the total error rate of STT-MRAM cache, which consists of retention failure, read disturbance, and write failure strongly depends on \textbf{a)} data patterns, \textbf{b)} cell contents, and \textbf{c)} data access times. 
			Fig. \ref{fig:motive} shows variations in an L2-cache access patterns for a set of workloads from SPEC CPU2006 benchmark suite~\cite{spec2006}.  
			Number of `1's read from block during the workload execution affect the read disturbance probability, number of ${0\rightarrow 1}$ and ${1\rightarrow 0}$ transitions affects the write failure probability, and the idle intervals excluding the intervals before a write operation affect the retention failure probability.
			The results illustrate significant variations for all of these parameters in different workloads.
			This observation indicates a strong workload-dependency of not only the rate of all error types, but also the contribution of each type of error in total error rate.  
			\subsection{Effects of Process Variations}	
			Similar to other memory technologies, STT-MRAM technology is affected by process variations. These variations are exacerbated by downscaling the technology node~\cite{imani2016approximate,kim2016exploration, kang2015reconfigurable, 12-EDCC-sun2012process}.
			PVs in the MTJ ferromagnetic layers and oxide barrier layer as well as in NMOS access transistor deviate the STT-MRAM parameters from their nominal values.
			Thermal stability factor ($\Delta$), read and write currents ($I_{read}$ and $I_{write}$), critical switching current ($I_{C_{0}}$), magnetic moment of MTJ free layer ($m$), and tunneling spin polarization ($p$) are among the major parameters affected by PVs.
			In the following, we explain how PVs affect the STT-MRAM cache reliability.

			As observed in (\ref{eq:4}), the occurrence probability of retention failure depends on a cell idle time and its thermal stability factor.
			Although the former is determined by only the workload behavior and architectural properties of the cache, the latter is highly under PVs.
			$\Delta$ of a cell is deviated from its nominal value by altering the cell barrier energy ($E_b$) affected by PVs.
			According to (\ref{eq:4}), the retention failure probability exponentially depends on the exponential of $\Delta$.
			Therefore, even a minor deviation in $\Delta$ of a cell excessively changes the retention failure probability.
			A recent study shows that the retention time of STT-MRAM cells in a memory cell can vary from a few minutes to several years\cite{nair2017vaet}.

			Similar to the retention failure probability, $\Delta$ exponentially affects an exponential function that generates the read disturbance probability, based on (\ref{eq:6}).
			Hence, a small decrease in $\Delta$ due to PVs extremely increases the read disturbance probability.
			Deviated from its nominal value, a small decrease in critical switching current ($I_{C_{0}}$) has the same effect as $\Delta$ deviation on the read disturbance probability.
			The effect of increase in read current ($I_{read}$) on increase in read disturbance probability is the same as the effect of decease in $\Delta$ and $I_{C_{0}}$.
			Therefore, all parameters in (\ref{eq:6}) that are under PVs, i.e., $\Delta$, $I_{C_{0}}$, and $I_{read}$, affect the read disturbance probability through the exponent of an exponential function.

			Write current ($I_{write}$), critical switching current ($I_{C_{0}}$), thermal stability factor ($\Delta$), magnetize moment of MTJ free layer ($m$), and tunneling spin polarization ($p$) are the parameters under PVs that affect write failure rate.
			According to (\ref{eq:6}), a rise in $I_{C_{0}}$, $\Delta$, $m$, and $p$ values due to PVs increase the write failure probability exponentially.
			On the other hand, this probability increases by decreasing $I_{write}$.

			In summary, as the cell contents, number of accesses, time interval between accesses, and PVs determine the cache error rate, we need to investigate the reliability of STT-MRAM caches in the presence of PVs and different workloads.
			$\Delta$, $I_{C_{0}}$, $I_{read}$, $I_{write}$, $m$, and $p$ are the main parameters in STT-MRAM cells affected by PVs and can significantly change the cache reliability in the following aspects:
			\begin{itemize}
			\item An increase (decrease) in $\Delta$ increases (reduces) the write failure rate in one hand and reduces (increases) the retention failure and read disturbance rates on the other hand.
			\item An increase (decrease) in $I_{C_{0}}$ increases (decreases) the write failure rate and decreases (increases) the read disturbance rate. The value of $I_{C_{0}}$ has no effect on retention failure rate.
			\item An increase (decrease) in $I_{read}$ increases (decreases) the read disturbance rate without affecting the other two error types.
			\item An increase (decrease) in $I_{write}$ decreases (increases) the write failure rate and has no effect on the other two error types.
			\item An increase (decrease) in $m$ and $p$ increases (decreases) the write failure rate and has no effect on the others.
			\end{itemize}

		\section{Proposed Formulations and Framework}	
			
			Retention failure, read disturbance, and write failure rates, which are affected by workloads and process variations cannot be measured through the cells physical characteristics. 
			In other words, the reliability of a STT-MRAM cache needs system-level investigations as well as circuit-level information. 
			The vulnerability of STT-MRAM cache to different error types requires a system-level exploration through a joint consideration of device-, circuit-, architecture-, and application-level effects. 
			Moreover, the reliability estimation of STT-MRAM cache requires the integration of the cache vulnerability to all error types, which their rates are differently and even oppositely affected from device- to application-level parameters.

This work presents a comprehensive reliability formulation of STT-MRAM cache by jointly considering all three error sources, i.e., write failure, read disturbance, and retention failure. In other words, our formulations are an integration of write failure, read disturbance, and retention failure rates. The total cache error rate under each of these error sources is derived from error probability of all cache blocks. Error probability of each cache block is the summation of all STT-MRAM cells of the block. The input parameters of all the proposed reliability formulas are the error rate of a single STT-MRAM cell. These parameters are derived practically from real experiments in several studies and the formulations are adjusted to match the experimental results.

			Early reliability exploration of STT-MRAM caches at design time is a necessity for not only confirming the applicability of STT-MRAM technology as a reliable alternative to SRAM technology, but also to design cost-efficient error-tolerant caches.
			In this section, we first present our formulations for the vulnerability of STT-MRAM caches to each error type.
			Then, the formulations for total cache vulnerability, which is the integration of cache vulnerability to all error types is presented.
			Finally, our framework, as an extension to gem5 full-system simulator~\cite{gem5}, for reliability exploration is detailed.

			\setcounter{secnumdepth}{4}
			\subsection{Cache Vulnerability Formulations}
			
			The vulnerability of STT-MRAM caches to error sources depends on \textbf{a)} the number and sequence of read and write accesses, \textbf{b)} the interval between accesses, and \textbf{c)} the content of stored and updated data value.
			 On the other hand, PVs affect the cache vulnerability by changing the STT-MRAM physical parameters and complicate the vulnerability formulations.
			 For the sake of simplicity and generality, we first formulate the cache vulnerability without considering PVs and then extend our formulations by including the PV effects, in the following subsections. 
			\vspace{5pt}
			\subsubsection{\textbf{Cache Vulnerability Formulation without PVs}}

			The probability of retention failure, read disturbance, and write failure for a single STT-MRAM cell was given in (\ref{eq:3}), (\ref{eq:5}), and (\ref{eq:6}), respectively, for a single interval/access.
			We extrapolate these equations to calculate error probabilities for an $N$-$bit$ cache block in a single interval/access and extend it for error probability of a block during workloads execution.
			Finally, the vulnerability of the cache is formulated using the error probability of all blocks.
\vspace{5pt}
\paragraph{\textbf{Retention Failure Probability}}

			The retention failure probability of an $N$-$bit$ cache block in an idle interval $t$ is according to (\ref{eq:7}).\vspace{-3pt}
			 \begin{equation}
			\begin{multlined}
			\label{eq:7}\hspace{-1cm}
			P_{RF,blk,t}=1- $$\prod_{i=0}^{N-1} (1- P_{RF, bit_{i},t})$$\\
			\shoveleft[0.5cm]{=1- $$\prod_{i=0}^{N-1} (exp({-t }\times {exp(-\Delta)})$$}\\
			\shoveleft[0.5cm]{=1- exp({-t }\times {exp(-\Delta)})^ {N}}\\
			\shoveleft[0.09cm]{=1- exp({-N\times t }\times {exp(-\Delta)})}
			\end{multlined}
			\end{equation}
			Where, $P_{RF,blk,t}$ is probability of retention failure (RF) for the block and $P_{RF,bit_i,t}$ is the retention failure probability for a single cell given in (\ref{eq:3}).
			A block is erroneous if a retention failure occurs in at least one out of its $N$ cells.
			In other words, a block is error-free if all of its cells are error-free.	
			Since the cache blocks are assumed to be accessed in the block granularity, the idle intervals are equal for all cells and the retention failure probability of all cells are the same.
			On the other hand, the probability of no retention failure occurrence in $N$ bits for interval $t$ is equivalent to that probability in a single bit for interval $N\times t$.
			Therefore, the retention failure probability is simplified according to (\ref{eq:7}).

\begin{figure}[t]
				\centering
				\includegraphics[width=1\linewidth]{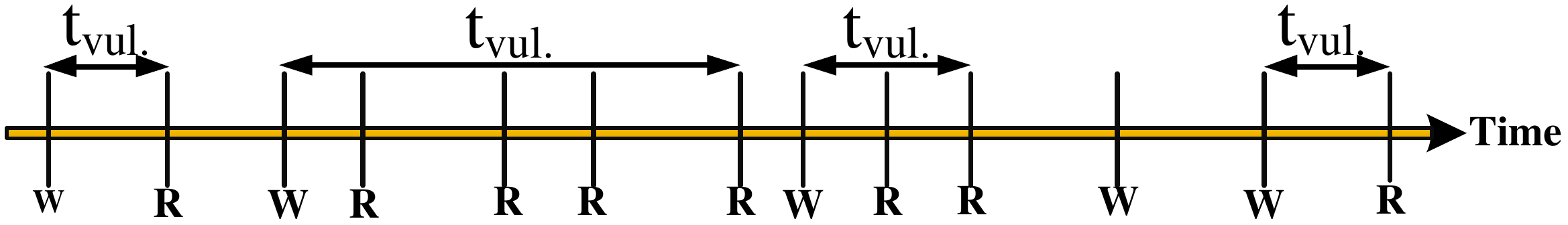}\vspace{-8pt}
				\caption{Vulnerable intervals ($t_{vul.}$) of cache blocks to retention failure occurrence.}\vspace{-5pt}
				\label{fig:4}
			\end{figure}

			A cache block experiences multiple idle intervals during a workload execution. 
			The retention failure probability for a block in total execution time is according to (\ref{eq:8}). \vspace{-3pt}
			 \begin{equation}
			\begin{multlined}
			\label{eq:8}
			P_{RF,blk,total}=1-$$\prod_{j=1}^{T} (1- P_{RF, blk,t_{j}})$$\\
			\shoveleft[2cm]{=1-$$\prod_{j=1}^{T} (exp({-N\times t_{j} }\times {exp(-\Delta)})$$}\\
			\shoveleft[2cm]{=1-exp({-N}\times $$\sum_{j=1}^{T} t_{j}\times {exp(-\Delta)})}
			\end{multlined}
			\end{equation}
			Where, $T$ is the total number of vulnerable idle intervals for the block and $P_{RF,blk,t_j}$ is the retention failure probability given in (\ref{eq:7}).	
			Vulnerable idle intervals of a block are idle intervals ended by a read access. 
			Retention failures occurred in idle intervals ended by a write access are overwritten. 
			Fig. \ref{fig:4} depicts vulnerable intervals of a block for a sample sequence of write/read accesses. 
			The retention failure rate of a block in total intervals is equivalent to that probability for a single interval that its duration is the summation of all intervals.

			Using (\ref{eq:8}), the retention failure probability of the cache in total execution time is calculated according to (\ref{eq:9}).	
			\begin{equation}
			\begin{multlined}
			\label{eq:9}
			P_{RF,cache}=1- $$\prod_{b=0}^{B-1} (1- P_{RF, blk_{b},total})$$\\
			\shoveleft[0.1cm]{=1- $$\prod_{b=0}^{B-1} (exp({-N}\times $$\sum_{j=1}^{T_{b}} t_{b,j}\times {exp(-\Delta)}))$$}\\
			\shoveleft[0.1cm]{=1- exp({-N}\times $$\sum_{b=0}^{B-1} $$\sum_{j=1}^{T_{b}} t_{b,j}\times {exp(-\Delta)})}
			\end{multlined}
			\end{equation}
			Where, $B$ is the total number of blocks in the cache and $T_b$ is total intervals of block $b$.
			The retention failure of the cache is equivalent to the retention failure of a single block that its interval is the summation of all intervals of all blocks.
\vspace{5pt}		
			\paragraph{\textbf{Read Disturbance Probability}}
			Based on the design time configurations, read disturbance error is probable in memory cells containing either logic value `1' or `0'.
			In this study, we assume that the direction of read current is the same as the current for writing `0', and only STT-MRAM cells containing `1' are vulnerable to read disturbance.
			It is noteworthy that the direction of read disturbance has no effect on our formulations.
			Using the read disturbance probability for a single cell in (\ref{eq:5}), we derive the read disturbance probability for an $N$-$bit$ cache block in a single read access, according to (\ref{eq:10}).
			\begin{equation}
			\begin{multlined}
			\label{eq:10}
			P_{RD,blk,RA}=1- $$\prod_{i=0}^{N-1}(1- P_{RD,bit_{i},RA})\\   
			\shoveleft[0.3cm]{=1- $$\prod_{i=0}^{N-1}(1- bit_{i}\times P_{RD,cell,RA})}\\
			\shoveleft[0.3cm]{=1-{(1- P_{RD,cell,RA}})^{n}} \\
			\shoveleft[0.005cm]{=1- (exp(\frac{-t_{read}}{\tau}\times exp(\frac{\Delta(I_{read}-I_{C_0})}{I_{C_0}})))^n}\vspace{2pt}
			\end{multlined}
			\end{equation}
			Where, $RD$ and $RA$ stands for Read Disturbance and Read Access, respectively.
			$P_{RD,bit_{i},RA}$ is the read disturbance probability of $bit_i$ in the block, which is equal to zero if $bit_i$= `0' and is equal to $P_{RD,cell,RA}$ given in (\ref{eq:5}), otherwise.
			Assuming $n$ bits out of the N-bit block contains `1', the read disturbance probability of a block in a single read access is equivalent to read disturbance probability in a cell containing `1' for $n$ read accesses.
			It is noteworthy that a block is erroneous if read disturbance occurs in at least one of its $N$ bits.

			A cache block is read for several times with different contents during a workload execution.
			The read disturbance probability for a block in total execution time is according to (\ref{eq:11}). 
			 \begin{equation}
			\begin{multlined}
			\label{eq:11}
			P_{RD,blk, total}=1- $$\prod_{i=1}^{R}(1- P_{RD,blk,RA_{i}})\\
			\shoveleft[0.001cm]{=1- $$\prod_{i=1}^{R}(1- P_{RD,cell,RA})^{n_i}} \\
			\shoveright[2cm]{=1- (exp(\frac{-t_{read}}{\tau}\times exp(\frac{\Delta(I_{read}-I_{C_0})}{I_{C_0}})))^{(\sum_{i=1}^{R} n_{i})}}
			\end{multlined}
			\end{equation}
			Where, $R$ is the total number of read accesses to the block, $P_{RD,blk,RA_{i}}$ is the read disturbance probability in a single read access $RA_{i}$, and $n_i$ is the number of `1's in the block for access $RA_{i}$.	
			Read disturbance probability for a block in $R$ accesses, in which $n_i$ bits contain `1' for ${i}\textsuperscript{th}$ access is equivalent to read disturbance probability in a cell containing `1' and is read for $\sum_{i=1}^{R} n_{i}$ times.

			 Using the read disturbance probability of a cache block, the read disturbance probability of the whole cache in total execution time is calculated according to (\ref{eq:12}). \vspace{-5pt}
			 \begin{equation}
			\begin{multlined}
			\label{eq:12}
			P_{RD,cache}=1- $$\prod_{b=0}^{B-1}(1- P_{RD,blk_{b},total})\\
			\shoveleft[1.6cm]{=1- $$\prod_{b=0}^{B-1}((1- P_{RD,cell,RA})^{n_i})^{(\sum_{i=1}^{R_{b}} n_{b,i})}} \\
			\shoveleft[1.6cm]{=1- (exp(\frac{-t_{read}}{\tau}\times}\\
			 \shoveleft[1.7cm]{exp(\frac{\Delta(I_{read}-I_{C_0})}{I_{C_0}})))^{(\sum_{b=0}^{B-1}\sum_{i=1}^{R_{b}} n_{b,i})}}
			\end{multlined}
			\end{equation}
			Where, $P_{RD,cache}$ is equivalent to read disturbance probability of a cell  containing `1' after reading it for the total number of `1's during execution time.
\vspace{5pt}
			\paragraph{\textbf{Write Failure Probability}}
			In a write operation, write failure is probable in STT-MRAM cells that their content requires flipping.
			On the other hand, the write failure probability for ${0\rightarrow1}$ transitions is different from that in ${1\rightarrow 0}$ transitions.
			Therefore, the write failure probability for a cache block in a single write operation depends on not only the number of switching required, but also the number of required switching in each direction.
			Having the write failure probability for a single cell in single write access, the write failure probability for a cache block in a write access is according to (\ref{eq:13}).\vspace{-5pt}
			\begin{equation}
			\begin{multlined}
			\label{eq:13}
			P_{WF,blk,WA}=1- $$\prod_{i=0}^{N-1}(1- P_{WF,bit_{i},WA}) \\
			\shoveleft[2cm]{=1- (1- P_{WF,cell_{0 \rightarrow1}})^{n_{0\rightarrow1}}\times}\\
			\shoveleft[2cm]{(1- P_{WF,cell_{1\rightarrow 0}})^{n_{1\rightarrow 0}}}
			\end{multlined}
			\end{equation}
			Where, $WF$ and $WA$ stand for Write Failure and Write Access, respectively, $P_{WF,bit_{i},WA}$ is the write failure probability in $bit_i$ of the block, and
			$P_{WF,cell_{0\rightarrow1}}$ and $P_{WF,cell_{1\rightarrow0}}$ are the write failure probability for ${0\rightarrow1}$ and ${1\rightarrow0}$ transitions, respectively.
			$n_{0\rightarrow 1}$ and $n_{1\rightarrow 0}$ are the number of ${0\rightarrow 1}$ and ${1\rightarrow 0}$ transitions, respectively.
			The write failure probability in a block is equivalent to this probability when trying to flip a cell containing `1' for $n_{1\rightarrow 0}$ times and a cell containing `0' for $n_{0\rightarrow 1}$ times.
				
			For a cache block written for several times during a workload execution, the total write failure probability can be derived from (\ref{eq:13}), as shown in (\ref{eq:14}).\vspace{-4pt}
			\begin{equation}
			\begin{multlined}
			\label{eq:14}
			P_{WF,blk,total}=1- $$\prod_{i=1}^{W}(1- P_{WF,blk,WA_{i}})\\
			\shoveleft[1cm]{=1- $$\prod_{i=1}^{W}((1- P_{WF,cell_{0\rightarrow1}})^{{(n_{0\rightarrow1})}_i}}\\
			\shoveleft[1.3cm]{\times(1- P_{WF,cell_{1\rightarrow0}})^{{(n_{1\rightarrow0})}_i})}\\
			\shoveleft[1cm]{=1-(1- P_{WF,cell_{0\rightarrow1}})^{(\sum_{i=1}^{W} {(n_{0\rightarrow1})}_i)}}\\
			\shoveleft[0.8cm]{\times (1- P_{WF,cell_{1\rightarrow0}})^{(\sum_{i=1}^{W} {(n_{1\rightarrow0})}_i)}}\vspace{-3pt}
			\end{multlined}
			\end{equation}
			Where, $W$ is the total number of write accesses to the block, $P_{WF,blk,WA_{i}}$ is the write failure probability in write access $i$, in which the number of required  $0\rightarrow 1$ and $1\rightarrow 0$ transitions are equal to ${(n_{0\rightarrow 1})}_i$ and ${(n_{1\rightarrow 0})}_i$, respectively.
			Write failure probability for a block during the total execution time is equivalent to this probability when trying to flip a cell containing `0' for ${\sum_{i=1}^{W} {(n_{0\rightarrow1})}_i}$ times and a cell containing `1' for ${\sum_{i=1}^{W} {(n_{1\rightarrow0})}_i}$ times.

			The write failure probability of the whole cache in total execution time is derived from this probability for the blocks, according to (\ref{eq:15}).\vspace{-10pt}
			
			\begin{equation}
			\begin{multlined}
			\label{eq:15}\vspace{6pt}
			P_{WF, cache}=1- $$\prod_{b=0}^{B-1}(1- P_{WF,blk_b,total})\\
			\shoveleft[1.5cm]{=1-$$\prod_{b=0}^{B-1}((1- P_{WF,cell_{0\rightarrow1}})^{(\sum_{i=1}^{W_b} {(n_{0\rightarrow1})}_{b,i})}}\vspace{6pt}\\
			\shoveleft[1.8cm]{\times (1- P_{WF,cell_{1 \rightarrow0}})^{(\sum_{i=1}^{W_b} {(n_{1\rightarrow0})}_{b,i})}}\vspace{6pt}\\
			\shoveleft[1.5cm]{=1-(1- P_{WF,cell_{0\rightarrow1}})^{(\sum_{b=0}^{B-1} \sum_{i=1}^{W_b} {(n_{0\rightarrow1})}_{b,i})}\vspace{6pt}\\
			\shoveleft[1.5cm]{\times (1- P_{WF,cell_{1\rightarrow0}})^{(\sum_{b=0}^{B-1} \sum_{i=1}^{W_b} {(n_{1\rightarrow0})}_{b,i})}}}
			\end{multlined}
			\end{equation}
			Where, $B$ is the number of blocks in the cache, $P_{WF,blk_b,total}$ is the write failure probability for block $b$ in total execution time, $W_b$ is the number of write accesses to block $b$, and ${(n_{0\rightarrow 1})}_{b,i}$ and ${(n_{1\rightarrow 0})}_{b,i}$ are the number of $0\rightarrow 1$ and $1\rightarrow 0$ transitions in write access $i$ to block $b$, respectively.
			Write failure probability in the whole cache is equivalent to this probability in a single cell when trying to flip its content from `1' to `0' and from `0' to `1' for $T_1$ and $T_2$ times, respectively, in which $T_1$ and $T_2$ are the total number of $0\rightarrow 1$ and $1\rightarrow 0$ transitions in the cell during workloads execution, respectively.
			
\vspace{5pt}
			\subsubsection{Cache Vulnerability Formulation with PV Effects}
			As observed in above formulations, the probability of errors in STT-MRAM caches depends on both workloads behavior and the physical characteristics of STT-MRAM cells.
			By considering PVs in our formulations, the error probability of different cache blocks and even that of different cells in a block differs for the same access pattern and cells content.
			In the following, we formulate the errors probability by including PV effects.
\vspace{5pt}
			\paragraph{\textbf{PV Affected Retention Failure Probability}}
			As observed in (\ref{eq:3}), the only parameter affected by PVs is the thermal stability factor ($\Delta$) in STT-MRAM cells.
			The retention failure probability of a cache block for a single idle interval $t$ is according to (\ref{eq:16}).\vspace{-1pt}
			 \begin{equation}
			\begin{multlined}
			\label{eq:16} \vspace{8pt}
			P_{RF,blk,t,PV}=1-$$\prod_{i=0}^{N-1}(1-P_{RF,bit_i,t,PV})\\	
			\shoveleft[1.5cm]{=1-$$\prod_{i=0}^{N-1} (exp({-t\times exp(-{\Delta}_i)}))$$}
			\end{multlined}
			\end{equation}
			Where, $P_{RF,bit_i,t,PV}$ is the retention failure probability of bit $i$ of the block in interval $t$ affected by PVs and $\Delta_i$ is the thermal stability factor of that bit, which can be different from that of other bits.

			The retention failure probability for a block in total execution time is according to (\ref{eq:17}).\vspace{-1pt}	
			 \begin{equation}
			\begin{multlined}
			\label{eq:17}\vspace{8pt}
			P_{RF,blk,total, PV}=1- $$\prod_{j=1}^{T} (1- P_{RF, blk,t_{j}, PV})$$\\
			\shoveleft[1cm]{=1- $$\prod_{j=1}^{T} ($$\prod_{i=1}^{N-1}(exp(-t_{j} \times exp(-{\Delta}_i)))$$)$$}\vspace{8pt}\\
			\shoveleft[0.7cm]{=1- $$\prod_{i=0}^{N-1}($$\sum_{j=1}^{T} t_{j}\times exp(exp(-{\Delta}_i)))}
			\end{multlined}
			\end{equation}
			Where, $T$ is the total number of idle intervals during the workload execution, $t_j$ is the duration of ${j}\textsuperscript{th}$ interval, and $P_{RF,blk,t_{j},PV}$ is the retention failure probability of the block in interval $t_j$.
			The retention failure probability of a cell in the block for $T$ intervals is equivalent to this probability for the cell in a single interval that its duration is the summation of all intervals.
			
			The retention failure probability for the total cache is given in (\ref{eq:18}).\vspace{-1pt}
			 \begin{equation}
			\begin{multlined}
			\label{eq:18}\vspace{8pt}
			P_{RF,cache,PV}= 1- $$\prod_{b=0}^{B-1} (1- P_{RF, blk_{b},total,PV})$$\\
			\shoveright[5cm]{=1- $$\prod_{b=0}^{B-1}($$\prod_{i=0}^{N-1}($$\sum_{j=1}^{T_{b}} t_{b,j}\times exp(exp(-{\Delta}_{b,i})))$$)$$}
			\end{multlined}
			\end{equation}
			Where, $P_{RF,blk_{b},total,PV}$ is the retention failure probability of block $b$ in total execution time, ${\Delta}_{b,i}$ is thermal stability factor for cell $i$ in block $b$, and $t_{b,i}$ is the interval $i$ of block $b$.
\vspace{5pt}
			\paragraph{\textbf{PV affected Read Disturbance Probability}}
			As observed in (\ref{eq:5}), PVs affect thermal stability factor ($\Delta$), read current ($I_{read}$), and critical switching current ($I_{C_{0}}$).
			The read disturbance probability of a cache block for a single read access is according to (\ref{eq:19}).	\vspace{-1pt}		
			\begin{equation}
			\begin{multlined}  
			\label{eq:19}\vspace{8pt}
			P_{RD,blk,RA,PV}= 1- $$\prod_{i=0}^{N-1}(1- P_{RD,bit_{i},RA,PV})\vspace{8pt}\\
			\shoveleft[0.06cm]{=1- $$\prod_{i=0}^{N-1}(exp(\frac{-t_{read}}{\tau}\times exp(\frac{{\Delta}_i ({(I_{read})}_{i}-{(I_{C_{0}})}_{i})}{{(I_{C_{0}})}_{i}})))^{bit_{i}}}\vspace{8pt}\\
			\shoveright[3cm]{=1- $$\prod_{i=0}^{N-1}(exp(\frac{-bit_{i}\times t_{read}}{\tau}\times exp(\frac{{\Delta}_i ({(I_{read})}_{i}-{(I_{C_{0}})}_{i})}{{(I_{C_{0}})}_{i}})))}
			\end{multlined}
			\end{equation}
			Where, $P_{RD,~bit_{i},RA,PV}$ is the read disturbance probability of bit $i$ in the block, $bit_i$ is value of that bit, and $\Delta_i$, (${I_{read}})_{i}$, and ${(I_{C_{0}})}_{i}$ are thermal stability factor, read current, and critical switching current for $bit_i$ for a block, respectively.
			Since the error is only probable in the bits containing `1', we put the value of bit $i$ ($bit_i$) in the exponent of read disturbance probability of that bit to eliminate the contribution of bits containing `0' in the read disturbance probability.

			The read disturbance probability of a cache block for total read accesses is given in (\ref{eq:20}).\vspace{-1pt}  
			 \begin{equation}
			\begin{multlined}
			\label{eq:20}\vspace{6pt}
			P_{RD,blk,total,PV}= 1- $$\prod_{j=0}^{R}(1- P_{RD,blk,RA_{j},PV})\vspace{6pt}\\
			\shoveleft[0.1cm]{=1- $$\prod_{j=1}^{R}($$\prod_{i=0}^{N-1}(exp(\frac{-bit_{j,i}\times t_{read}}{\tau}\times }\vspace{6pt}\\
			\shoveleft[0.5cm]{exp(\frac{{\Delta}_i ({(I_{read})}_{i}-{(I_{C_{0}})}_{i})}{{(I_{C_{0}})}_{i}}))))}\vspace{6pt}\\
			\shoveleft[0.1cm]{=1-$$\prod_{i=0}^{N-1}(exp(\frac{-\sum_{j=1}^{R} bit_{i,j}\times t_{read}}{\tau}}\vspace{6pt}\\
			\shoveright[2cm]{\times exp(\frac{{\Delta}_i ({(I_{read})}_{i}-{(I_{C_{0}})}_{i})}{{(I_{C_{0}})}_{i}})))}
			\end{multlined}
			\end{equation}
			Where, $R$ is the total number of read accesses, $P_{RD,blk,RA_{j},PV}$ is the read disturbance probability in read access $j$, and $bit_{j,i}$ is the value of bit $i$ of the block in read access $j$. 
			
			Finally, we calculate the read disturbance probability for the whole cache using (\ref{eq:21}). 	\vspace{-1pt}	
			\begin{equation}
			\begin{multlined}
			\label{eq:21}\vspace{6pt}
			P_{RD,cache,PV}=1-$$\prod_{b=0}^{B-1}(1- P_{RD,blk_{b},total,PV})\vspace{6pt}\\
			\shoveleft[0.7cm]{=1- $$\prod_{b=0}^{B-1}($$\prod_{i=0}^{N-1}(exp(\frac{-\sum_{j=1}^{R} bit_{b,i,j}\times t_{read}}{\tau}\times}\vspace{6pt}\\
			\shoveleft[1cm]{exp(\frac{{\Delta}_{b,i} ({(I_{read})}_{b,i}-{(I_{C_{0}})}_{b,i})}{{(I_{C_{0}})}_{b,i}}))))}
			\end{multlined}
			\end{equation}
			Where, $P_{RD,blk_{b},total,PV}$ is the read disturbance probability of block $b$ in total execution time. 
\vspace{5pt}
			\paragraph{\textbf{PV Affected Write Failure Probability}}
			As observed in (\ref{eq:6}), PVs affect the write failure probability by changing several parameters to STT-MRAM cells.
			The write failure probability in a cache block for single write access is according to (\ref{eq:22}). \vspace{-10pt}
			
			\begin{equation}
			\begin{multlined}
			\label{eq:22}\vspace{6pt}
			P_{WF,blk,WA,PV}=1- $$\prod_{i=0}^{N-1}(1- P_{WF,bit_{i},WA,PV})\vspace{6pt} \\
			\shoveleft[0.1cm]{=1- $$\prod_{i=1}^{N-1}((1- P_{WF,{(cell_i)}\textsubscript{$0\rightarrow1$}})^{{(bit_{i})}\textsubscript{$0\rightarrow1$}}}\vspace{6pt}\\
			\shoveright[1.1cm]{\times(1- P_{WF,{(cell_i)}\textsubscript{$1\rightarrow0$}})^{{(bit_{i})}\textsubscript{$1\rightarrow0$}})}
			\end{multlined}
			\end{equation}
			Where, $P_{WF,bit_{i},WA,PV}$ is the write failure probability for bit $i$ in the block, which its value depends on the direction of transition required for that bit in this write operation.
			$P_{WF,{(cell_{i})}\textsubscript{$0\rightarrow 1$}}$ and $P_{WF,{(cell_{i})}\textsubscript{$1\rightarrow0$}}$ are the write failure probability in $cell_i$ for ${0\rightarrow1}$ and ${1\rightarrow0}$ transitions, respectively. 
			The value of ${(bit_i)}\textsubscript{$0\rightarrow1$}$ or ${(bit_i)}\textsubscript{$1\rightarrow0$}$ is one if this bit needs a ${0\rightarrow1}$ or ${1\rightarrow0}$ transition, respectively.
			
			The write failure probability of a cache block for all write accesses in total workload execution is according to (\ref{eq:23}). \vspace{-1pt}
			\begin{equation}
			\begin{multlined}
			\label{eq:23}\vspace{6pt}
			P_{WF,blk,total,PV}=1- $$\prod_{j=1}^{W}(1- P_{WF,blk,WA_{j},PV})\vspace{6pt}\\
			\shoveleft[0.1cm]{=1- $$\prod_{j=1}^{W}($$\prod_{i=0}^{N-1}((1- P_{WF,(cell_{i})\textsubscript{$0\rightarrow1$}}))^{{(bit_{j,i})}\textsubscript{$0\rightarrow1$}}}\vspace{6pt}\\
			\shoveleft[0.5cm]{\times(1- P_{WF,(cell_{i})\textsubscript{$1\rightarrow0$}}))^{{(bit_{j,i})}\textsubscript{$1\rightarrow0$}})}\\
			\shoveleft[0.1cm]{=1- $$\prod_{i=0}^{N-1}((1- P_{WF,{(cell_{i})}\textsubscript{$0\rightarrow1$}})^{(\sum_{j=1}^{W} {(bit_{i,j})}\textsubscript{$0\rightarrow1$})}}\vspace{6pt}\\
			\shoveleft[0.1cm]{\times (1- P_{WF,{(cell_{i})}\textsubscript{$1\rightarrow0$}})^{(\sum_{j=1}^{W} {(bit_{i,j})}\textsubscript{$1\rightarrow0$})})}
			\end{multlined}
			\end{equation}
			Where, $W$ is the total number of write accesses to the block, $P_{WF,blk,WA_i,PV}$ is the write failure probability in write access $i$ given in (\ref{eq:22}), and ${(bit_{j,i})}\textsubscript{$0\rightarrow1$}$ and ${(bit_{j,i})}\textsubscript{$1\rightarrow0$}$ are one if bit $i$ requires a ${0\rightarrow1}$ or ${1\rightarrow0}$ transition in write access $j$, respectively.

			Finally, we formulate write failure probability in the whole cache in total execution time according to (\ref{eq:24}).	\vspace{-1pt}		
			\begin{equation}
			\begin{multlined}
			\label{eq:24}\vspace{6pt}
			P_{WF,cache,PV}= 1- $$\prod_{b=0}^{B-1}(1- P_{WF,blk_{b},total, PV})\vspace{6pt}\\
			\shoveleft[0.001cm]{=1- $$\prod_{b=0}^{B-1}$$\prod_{i=0}^{N-1}((1- P_{WF,{(cell_{b,i})}\textsubscript{$0\rightarrow1$}})^{(\sum_{j=1}^{W_{b}} {{(bit_{b,i,j})}\textsubscript{$0\rightarrow1$}})}}\vspace{6pt}\\
			\shoveright[1.2cm]{\times (1- P_{WF,{(cell_{b,i})}\textsubscript{$1\rightarrow0$}})^{(\sum_{j=1}^{W_{b}} {{(bit_{b,i,j})}\textsubscript{$1\rightarrow0$}})})}
			\end{multlined}
			\end{equation}
			Where, $P_{WF,blk_{b},total,PV}$ is the write failure probability of block $b$, $P_{WF,{(cell_{b,i})}\textsubscript{$0\rightarrow1$}}$ and $P_{WF,{(cell_{b,i})}\textsubscript{$1\rightarrow0$}}$ are the write failure probabilities in cell $i$ of block $b$ in ${0\rightarrow1}$ and ${1\rightarrow0}$ transitions, respectively, and the value of ${(bit_{b,i,j})}\textsubscript{$0\rightarrow1$}$ (${(bit_{b,i,j})}\textsubscript{$1\rightarrow0$}$) is one if bit $i$ of block $b$ needs ${0\rightarrow1}$ (${1\rightarrow0}$) transition in write access $j$.

			\subsection{Total Cache Failure Probability}
			
			Earlier, we formulated the occurrence probability of each error type in an STT-MRAM cache during workload execution time.
			To measure the cache reliability, we need to calculate the total cache failure probability considering all sources of errors.
			A STT-MRAM cache block is always vulnerable to \textit{only} one source of errors at any instant of time during the block lifetime.
			A block either is being read and vulnerable to read disturbance, is being written and vulnerable to write failure, or is idle and vulnerable to retention failure.
			On the other hand, for a read/write access to a block, only the accessed block is vulnerable to read disturbance/write failure, while all other blocks are vulnerable to retention failure, simultaneously.
			Therefore, for a STT-MRAM cache with $B$ blocks, in any instant of time, either all blocks are only vulnerable to retention failure or one block is vulnerable to read disturbance/write failure while the remaining $B-1$ blocks are vulnerable to retention failure.

			Calculating the total cache failure probability requires integrating all probable errors occurred in the cache originated from three exclusive sources with overlapping occurrence time.
			Retention failure is calculated in unit of time, write failure is formulated based on the number of write accesses, and read disturbance is a function of the number of read accesses.
			Integration of all error types requires transforming their probability into a same unit of measurement.
			To this aim, we reformulate the probability of retention failure, read disturbance, and write failure in unit of time.
			Using the total retention failure, read disturbance, and write failure probabilities of the cache given in (\ref{eq:9}), (\ref{eq:12}), and (\ref{eq:15}), respectively, the reliability in accordance with retention failure, read disturbance, and write failure are derived according to (\ref{eq:25}), (\ref{eq:26}), and (\ref{eq:27}), respectively.\vspace{-1pt}	 
			 \begin{equation}
			\label{eq:25}
			{R_{RF}(t)=exp(-N \times \frac{\sum_{b=0}^{B-1} \sum_{i=1}^{T_b}  t_{b,i}}{t_{exe}}\times exp(-\Delta)) }
			\end{equation}

			\begin{equation}
			\begin{multlined}
			\label{eq:26}
			{R_{RD}(t)=exp(\frac{-t_{read}}{\tau} \times} {exp(\frac{\Delta (I_{read}-{I_{C_0}})}{I_{C_0}}))^{\frac {(\sum_{b=0}^{B-1} \sum_{i=1}^{R_b}n_{b,i})}{t_{exe}}}}
			\end{multlined}
			\end{equation}

			\begin{equation}
			\begin{multlined}
			\label{eq:27}
			{R_{WF}(t)=(1-P_{WF, cell_{1\rightarrow0}})^{\frac{(\sum_{b=0}^{B-1} \sum_{i=1}^{W_b} {(n_{1\rightarrow0})\textsubscript {b,i}})}{t_{exe}}}\times}  \\
			\shoveleft[0.85cm]{(1-P_{WF, cell_{0\rightarrow1}})^{\frac{(\sum_{b=0}^{B-1} \sum_{i=1}^{W_b} {(n_{0\rightarrow1})\textsubscript {b,i}})}{t_{exe}}}}
			\end{multlined}
			\end{equation}
			Where, $R_{RF}(t)$, $R_{RD}(t)$, and $R_{WF}(t)$ are the occurrence probability of retention failure, read disturbance, and write failure per unit of time, respectively, and $t_{exe}$ is the total execution time for which the vulnerable interval $t_{b,i}$, total number of reads for each block ($R_b$), number of `1's in a cache read access ($n_{b,i}$), total number of writes for a cache block ($W_b$), and total number of ${0\rightarrow1}$ and ${1\rightarrow0}$ transitions for each write (${(n_{0\rightarrow1})}\textsubscript{$b,i$}$ and ${(n_{1\rightarrow0})}\textsubscript{$b,i$}$) are calculated. 
			The final step is to calculate the total cache failure probability per unit of time by integrating the above three reliability equations, according to (\ref{eq:28}).	\vspace{-10pt}

			\begin{equation}
			\label{eq:28}
			P_{TF/t}=1- (R_{RF}(t)\times R_{RD}(t)\times R_{WF}(t))
			\end{equation}

			The effects of PVs is not included in the above formulations.
			To calculate the cache reliability in terms of retention, read disturbance, and write failure considering PVs, we simply use PV affected probability given in (\ref{eq:18}), (\ref{eq:21}), and (\ref{eq:24}), instead of probabilities in (\ref{eq:9}), (\ref{eq:12}), and (\ref{eq:15}), respectively, for calculating the total cache failure probability. 

			The parameter $P_{TF}/t$ contains all factors that affect the reliability of an STT-MRAM cache.
			It integrates all conflicting physical parameters as well as cache access patterns and data contents.
			$P_{TF}/t$ is a key parameter for early reliability exploration of STT-MRAM caches in various scenarios of different device-level to system-level configurations.

An error in a cache block, regardless of its source (retention failure, write failure, and read disturbance) is manifested through a read operation. Therefore, all three sources of error contribute in the probability of reading an erroneous cache block. A block is error-free on a read operation if \textit{none} of the three types of errors occurs. In our evaluations, the occurrence probability of error in a block is calculated on each read operation on that block by integrating the occurrence probability of retention failure during its latest idle interval, write failure in the most recent write operation, and read disturbance in the current read operation. Therefore, while both retention and write failures are coupled with a read disturbance in their subsequent read operation, each of these three error sources has its own contribution in total error probability of a block. In this regards, we \textit{not}~\textit{only} report the total error rate of cache blocks, but also the \textit{breakdown} of the three error sources.

				\subsection{Architectural Aspects of Proposed Framework}
					
				An early reliability exploration considering device- and circuit-level  parameters affected by PVs as well as architecture- and system-level configurations and behaviors requires two main steps.
				First, modeling an STT-MRAM cache considering STT-MRAM cells physical characteristics and circuit-level cache configurations.
				Second, calculating the reliability metrics, i.e., $R_{RF}(t)$, $R_{RD}(t)$, $R_{WF}(t)$, and $P_{TF/t}$, using the proposed formulations for the target application.

				To this aim, we develop a framework that includes all necessary components for reliability exploration and integrate it to gem5 full-system simulator \cite{gem5}.
				This framework keeps track of runtime cache accesses to extract the required information for cache block, including content of each block cell at read and write accesses, content of updated data at write accesses, and the intervals between accesses.
				The framework imports the physical- and circuit-level parameters as well as different PVs models for STT-MRAM cells.

\begin{figure}[t]
				\centering
				\includegraphics[width=1\linewidth]{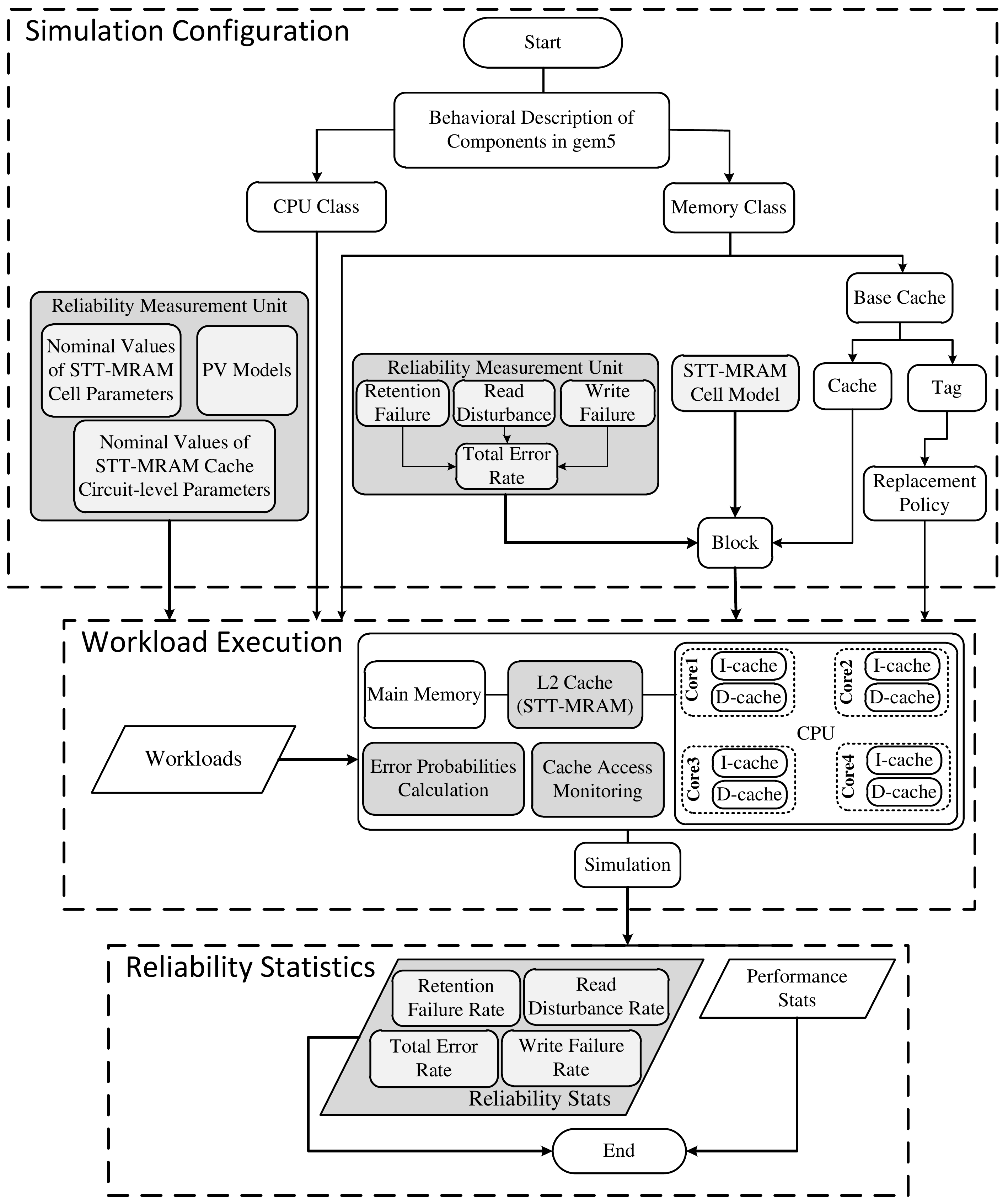}\vspace{-4pt}
				\caption{Flow of integrating the proposed framework with gem5 simulator.}\vspace{-15pt}
				\label{fig:6}
			\end{figure}

				Using the mentioned inputs, the framework calculates the rate of each error per cache block and the whole cache as well as the total error rate per block and the whole cache.
				This framework is also configurable to report the instantaneous error rates and the rates for any time frame.
				Fig. \ref{fig:6} shows an abstract view of gem5 simulator and the components of the reliability exploration framework appended to gem5.
				The boxes in gray are our framework components.

				The physical characteristics of STT-MRAM cells including both nominal and PV affected values as well as the circuit-level configurations of the cache for read/write accesses are included to gem5 at configuration time and the simulator is rebuilt for the simulation.
				At runtime, the cache blocks and access requests are monitored and the error probabilities and rates are calculated based on their corresponding formulas.
				By the end of simulation, the reliability statistics are reported besides the performance statistics.

			\section{Evaluation Methodology}

			In this section, we experimentally illustrate the dependency of STT-MRAM cache reliability to the workloads behaviors and quantitatively show the extreme error rate variation across different workloads.
			In addition, the effects of PVs on the cache reliability are investigated and the observations are discussed. 
			We integrate our proposed framework with gem5 full-system simulator~\cite{gem5}.
			A quad-core processor is modeled, which consists of dedicated instruction- and data- cache per core and a shared L2-cache.
			The L2-cache blocks are made of STT-MRAM cells that their detailed parameters are given in Table I and we conduct our reliability evaluations based on this cache configuration.
			These parameters are reported in 32nm technology node. To take the PV effects into account, we consider the Gaussian distribution for each PV affected parameter that its mean value is according to Table \ref{table:0}. 
			The standard deviation for all PV affected parameters is 0.05 of their nominal value, according to \cite{Eken-DAC-2016}.
			The system configuration details are given in Table \ref{table:1}.

			Workloads are 18 multi-programs generated by combining the workloads in SPEC CPU2006 benchmark suite~\cite{spec2006}.
			The details of these combinations are shown in Table \ref{table:2}.  
			The first one billion instructions in each workload execution are skipped for warm-up phase and the results are reported for the next four billion instructions.
			In the following, we first evaluate the effect of workloads on each error probability, separately, with no PV consideration.
			Then, the effects of PVs on the probability of each error type are investigated. 
			In the last subsection, we present the total cache failure probability for all workloads \textit{without} and \textit{with} PVs consideration as well as the contribution of each error type in the total failure probability.
			The failure probabilities are reported in microsecond as a unit of time.

			\begin{table}[t]
				\centering\vspace{-1pt}
				\caption{STT-MRAM cell parameters and their values}\vspace{-10pt}
				\includegraphics[width=1\linewidth]{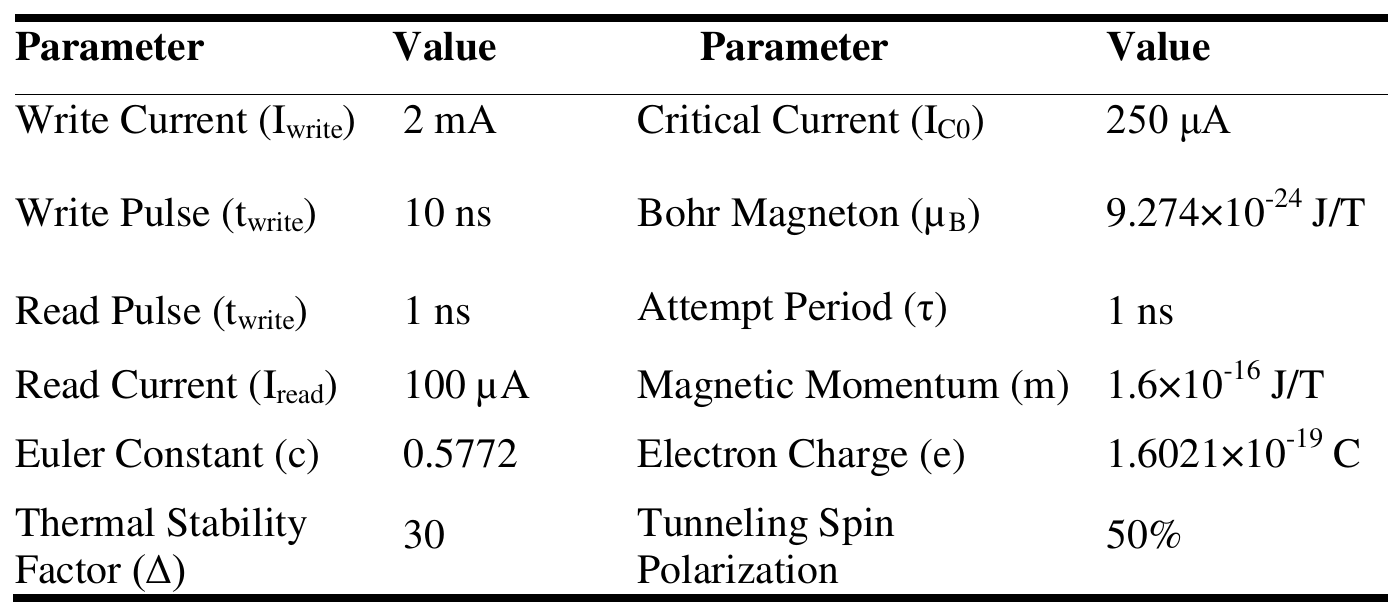}
				\label{table:0}\vspace*{-15pt}
			\end{table}

	\begin{table}[t]
				\centering\vspace{-1pt}
				\caption{Details of processor and caches configuration}\vspace{-10pt}
				\includegraphics[width=0.9\linewidth]{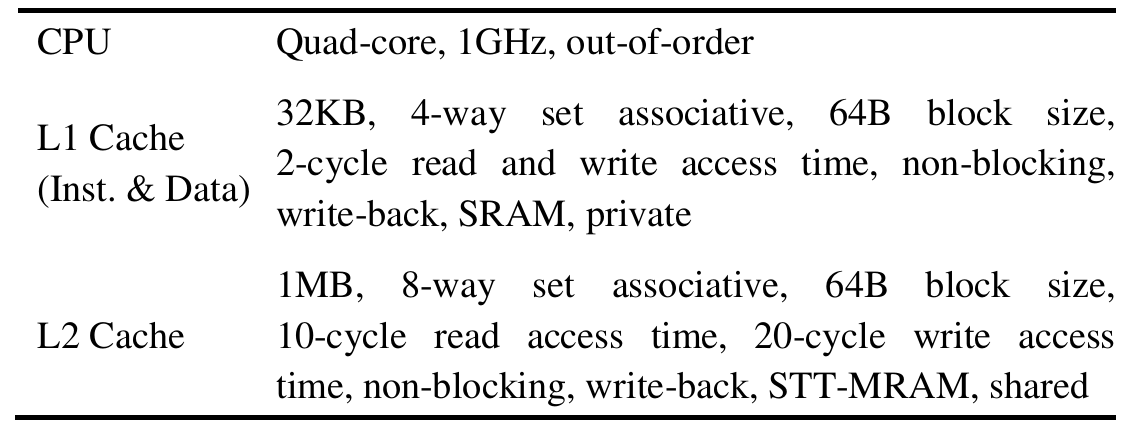}
				\label{table:1}\vspace*{-10pt}
			\end{table}

			\begin{table}[t]
				\centering
				\caption{Combinations of multi-programmed workloads for quad-core processor}\vspace{-10pt}
				\includegraphics[width=0.85\linewidth]{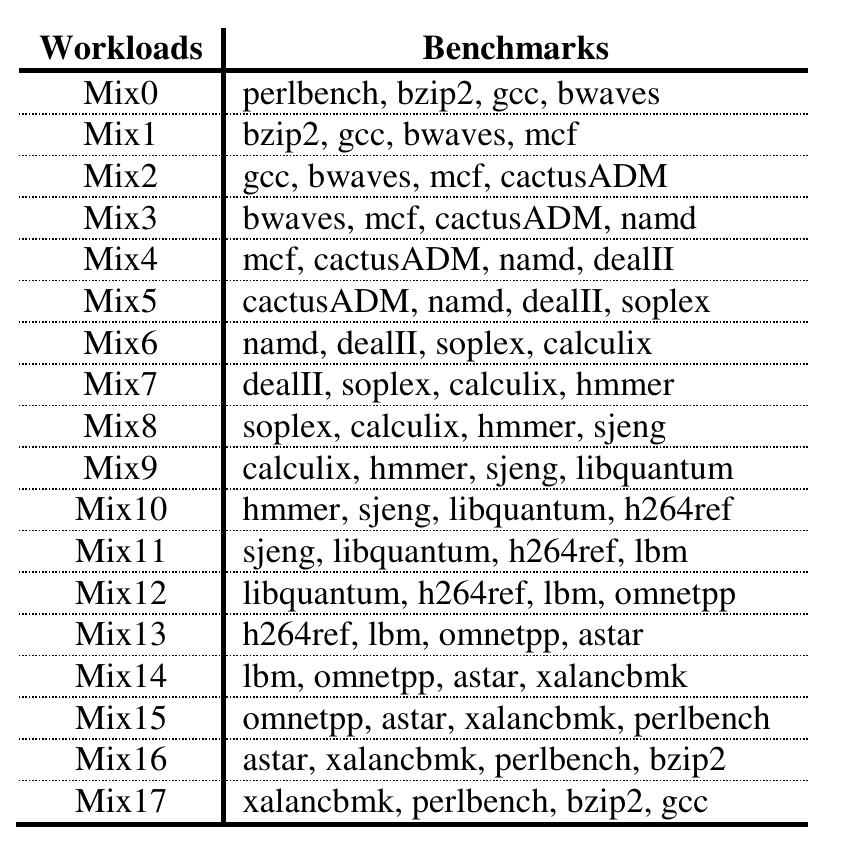}
				\label{table:2}\vspace*{-15pt}
			\end{table}

		\subsection{Workload Effects on Error Probabilities}

			The effects of workloads on error rates are twofold: \textbf{a)} the pattern of cache requests, and \textbf{b)} the content of stored and updated data blocks.
			The rate of read disturbance depends on both the number of read accesses and the content of the read blocks; the rate of write failure depends on both the number of write accesses and the previous and current content of written blocks; and finally, the rate of retention failure depends on the number of read/write accesses, the intervals between accesses to blocks, and the combination of read and write accesses to a block.

			For a deep analysis and a more detailed investigation, we present both the error probabilities considering all effects of workloads (access sequences and blocks contents) and the worst case error probability based on the cache access patterns.
			In the worst case error probability for read disturbance and write failure, it is assumed that all block cells are vulnerable regardless of their content.
			Therefore, the effect of data patterns is excluded in this case.
			In the worst case error probability for retention failure, all idle intervals are included, whereas the intervals between a read and its subsequent write access are excluded when considering only vulnerable intervals.
			Reporting error probabilities in both mentioned schemes unfolds the breakdown of workloads effects on cache reliability.

			\begin{figure}[t]
				\centering
				\includegraphics[width=1\linewidth]{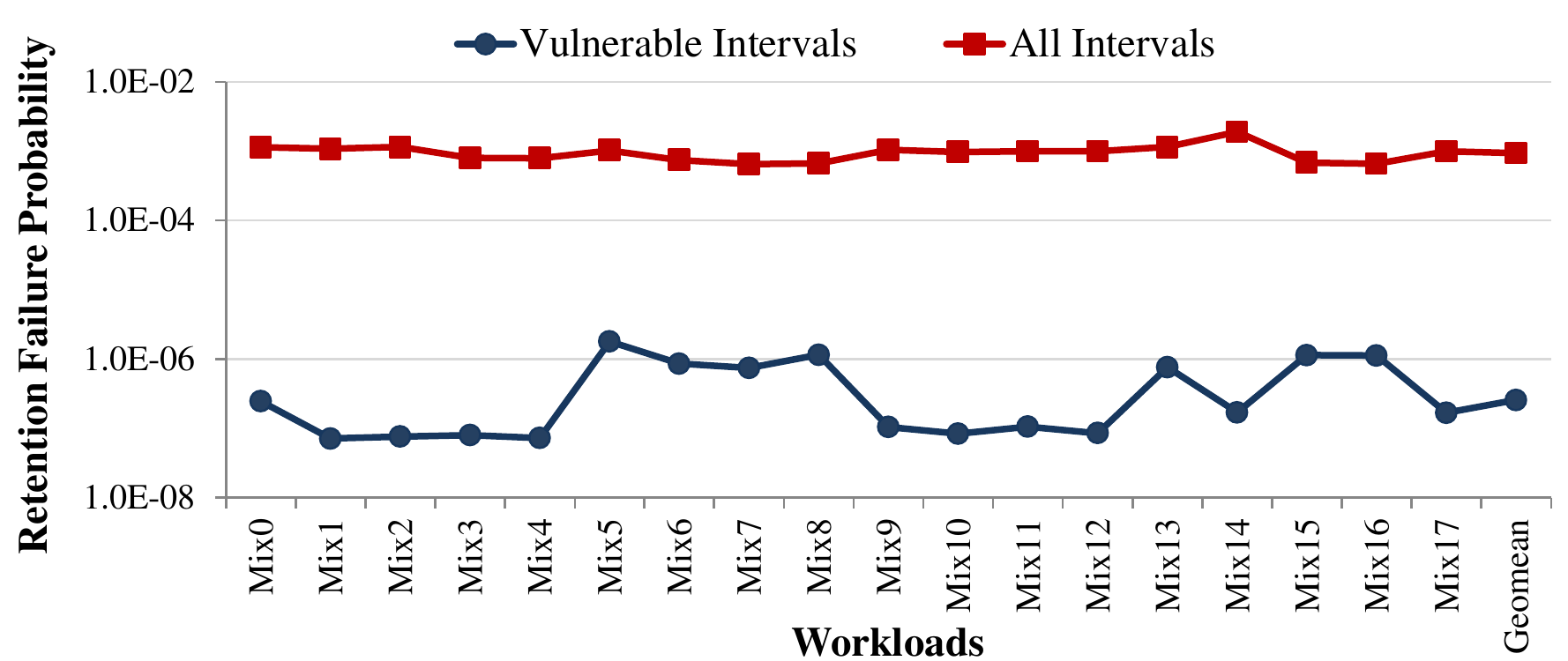}\vspace{-8pt}
				\caption{Effects of workloads behavior on retention failure probability with/without excluding the intervals that errors are masked.}\vspace{-10pt}
				\label{fig:7}
			\end{figure}

			Fig. \ref{fig:7} shows the retention failure probability for all workloads.
			The average error probability varies from 7.1$\times$$10^{\text{-8}}$ in $Mix1$ workload to 1.8$\times$$10^{\text{-6}}$ in $Mix5$ workload, indicating 25.4x variation between the error probabilities of different workloads.
			The average probability of retention failure is 2.6$\times$$10^{\text{-7}}$.
			By including the intervals between the read and its subsequent write access, the error probability is increased by more than 3 orders of magnitude, on average.
			The variation between error probabilities of different workloads in the worst case is due to various execution times.
			This variation is much lower than the time considering only vulnerable intervals.

			\begin{figure}[t]\vspace{-13pt}
				\centering
				\includegraphics[width=1\linewidth]{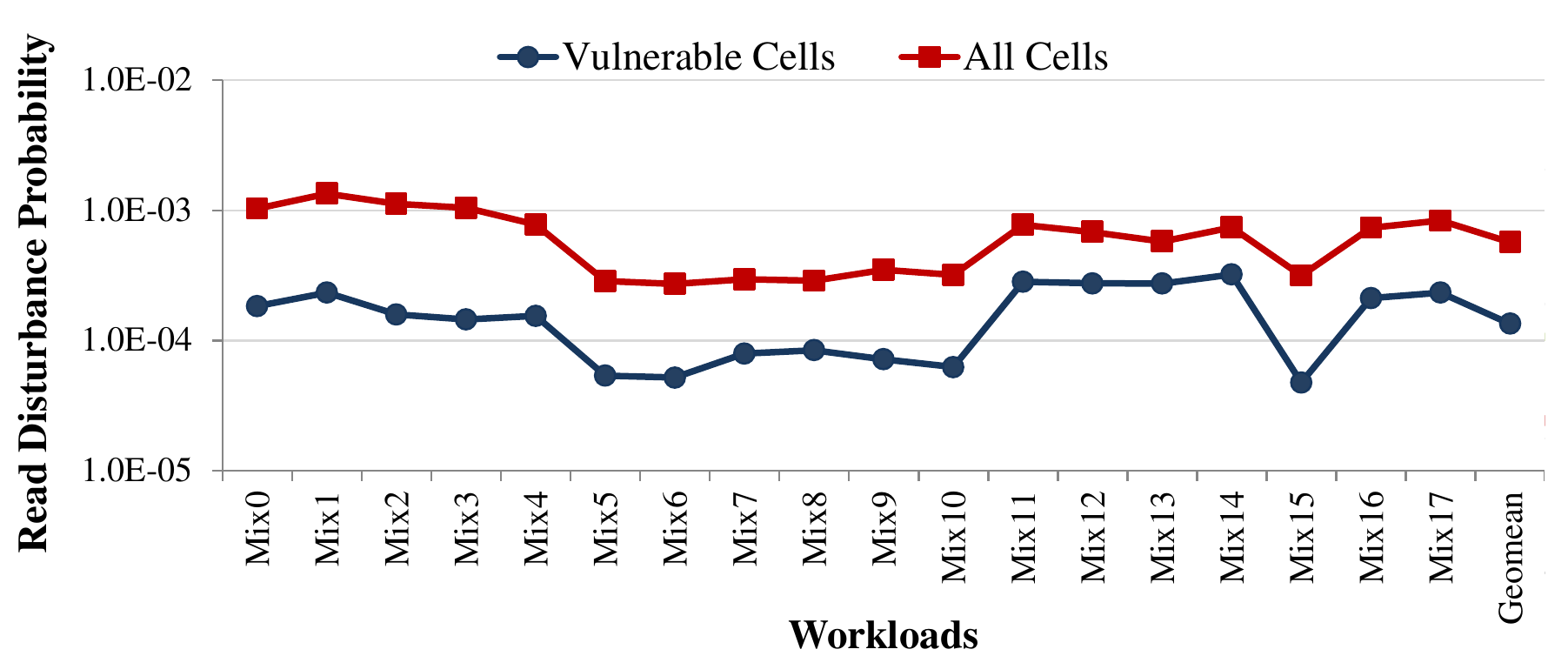}\vspace{-8pt}
				\caption{Effects of workloads behavior on read disturbance probability with/without considering the cache blocks content (content-dependent vs. worst case error rate).}\vspace{-10pt}
				\label{fig:9-rate}
			\end{figure}

			\begin{figure}[t]
				\centering
				\includegraphics[width=1\linewidth]{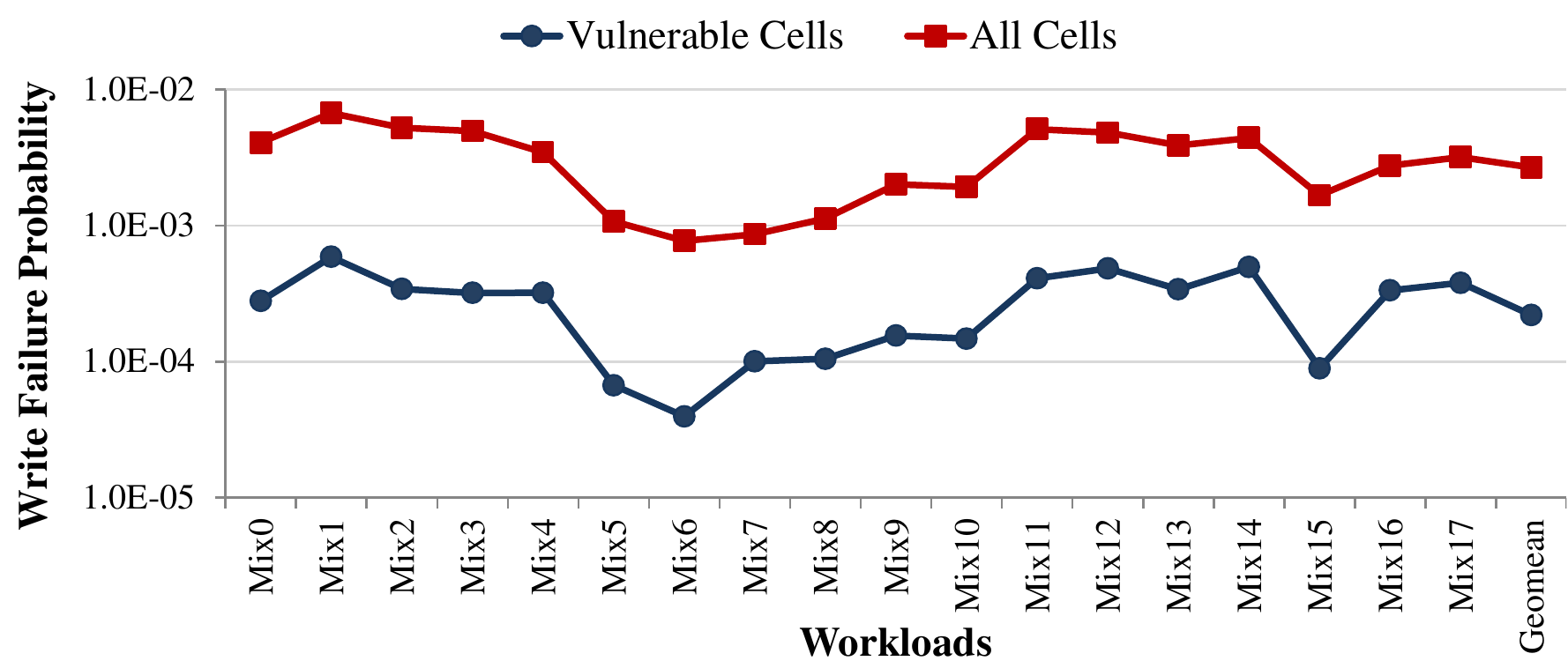}\vspace{-8pt}%
				\caption{Effects of workloads behavior on write failure probability with/without considering the required transitions on a write access (content-dependent vs. worst case error rate).}\vspace{-12pt}
				\label{fig:7write}
			\end{figure}

			The read disturbance probability for all workloads is shown in Fig. \ref{fig:9-rate}. 
			A large variation is observed in both content-dependent and worst case error probabilities.
			The content-dependent error probability varies from 4.7$\times$$10^{\text{-5}}$ in $Mix15$ workload to 3.2$\times$$10^{\text{-4}}$ in $Mix14$ workload, indicating 6.8x variations, while the worst case error probability varies from 2.7$\times$$10^{\text{-4}}$ in $Mix6$ workload to 1.3$\times$$10^{\text{-3}}$ in $Mix1$ workload, indicating 4.8x variation.
			From another perspective, the content of the data not only causes a large variation in error probabilities, as observed in  content-dependent error probabilities, but also significantly reduces the error probability, as the large gap between content-dependent and worst case probabilities confirms this fact.
			On average, the content-dependent error probability is by 4.2x lower than the worst case error probability, which is due to large number of `0's in cache blocks.
			However, in some workload, e.g., $Mix12$, $Mix13$, and $Mix14$, the content-dependent error probability is close to that in the worst case.   
		
			\begin{figure}[t]\vspace{-10pt}
				\centering
				\includegraphics[width=1\linewidth]{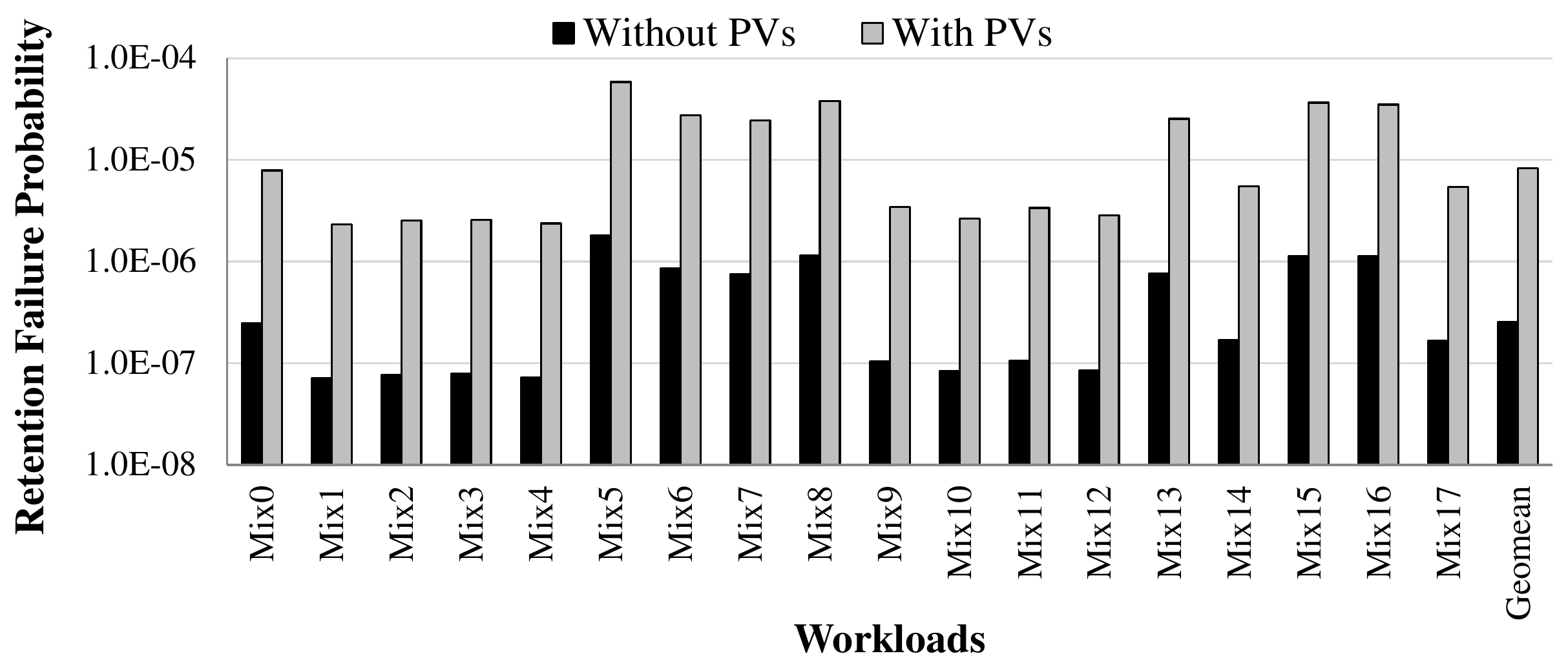}\vspace{-8pt}%
				\caption{Retention failure probability without/with considering PV effects.}\vspace{-5pt}
				\label{fig:9}
			\end{figure}
			
			Fig. \ref{fig:7write} depicts the write failure probabilities for the workloads.
			Assuming all block cells require a ${0\rightarrow1}$ transition, the worst case error probability is 2.7$\times$$10^{\text{-3}}$, on average.
			This value varies from 7.7$\times$$10^{\text{-4}}$ in $Mix6$ workload to 6.7$\times$$10^{\text{-3}}$ in $Mix1$ workload.
			Including the data content, the average error probability varies from 3.9$\times$$10^{\text{-5}}$ in $Mix6$ workload to 5.9$\times$$10^{\text{-4}}$ in $Mix1$ workload, which indicates a 15.1x variation.

	\begin{figure}[t]
				\centering
				\includegraphics[width=1\linewidth]{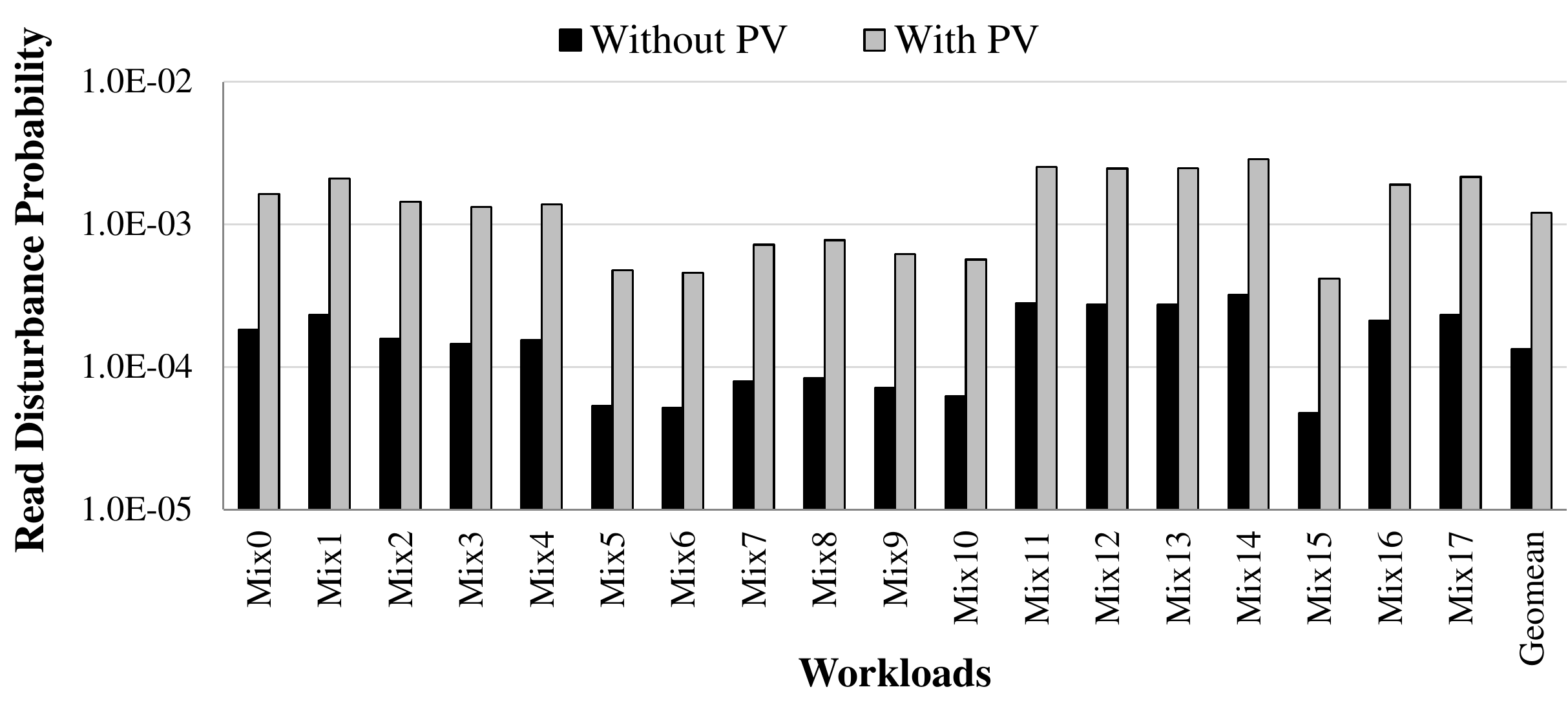}\vspace{-10pt}%
				\caption{Read disturbance probability without/with considering PV effects.}\vspace{-13pt}
				\label{fig:10}
			\end{figure}

			\subsection{PV Effects on Error Probabilities}

			\begin{figure}[b]\vspace{-10pt}
				\centering
				\includegraphics[width=1\linewidth]{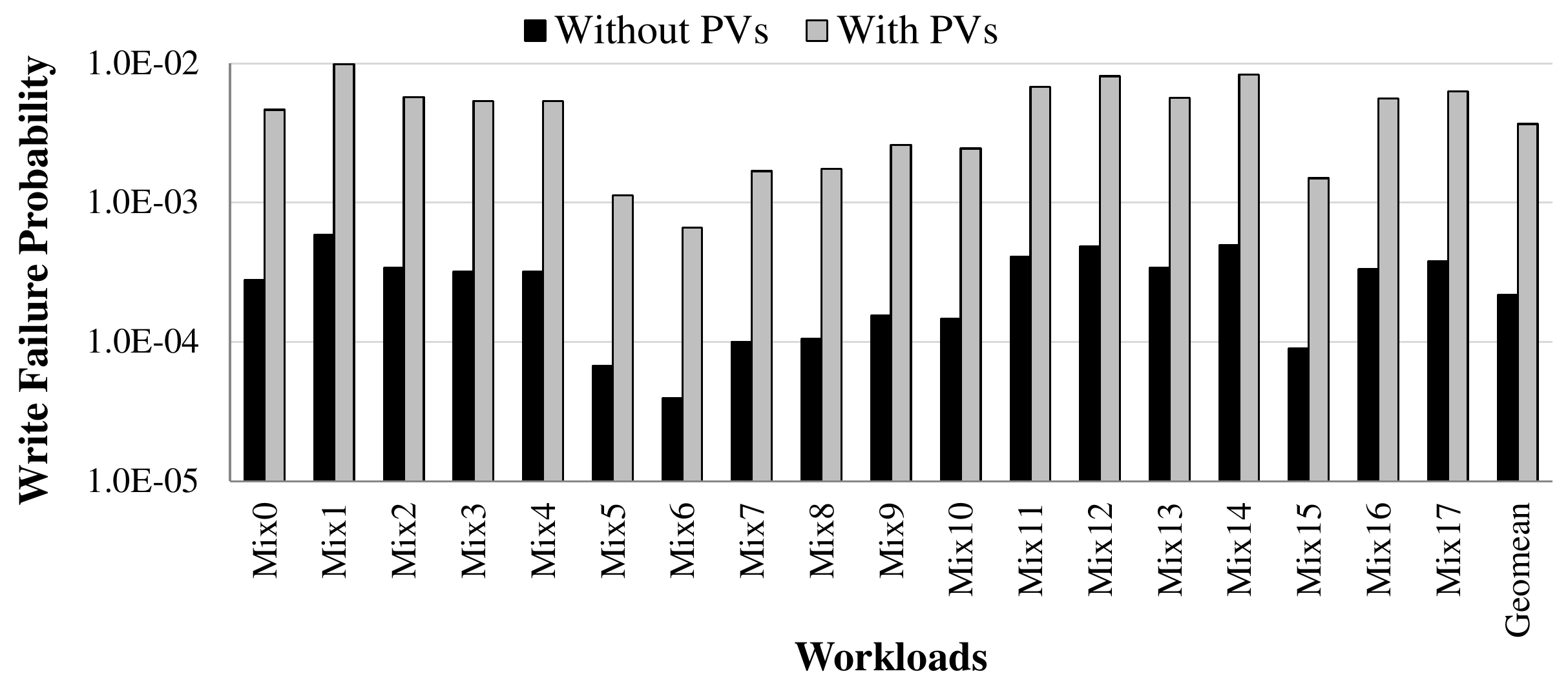}\vspace{-8pt}%
				\caption{Write failure probability without/with considering PV effects.}\vspace{-7pt}
				\label{fig:11}
			\end{figure}
			
			The effect of PVs on the vulnerability of a single cell can easily be analyzed based on the deviations in the physical cell parameters.
			However, due to dependency of error probabilities to the workloads behavior, the effect of PVs on the cache error probabilities needs system-level exploration.
			In the following, we report the probability of cache error type in the presence of PVs.

			Fig. \ref{fig:9} shows the retention failure probability for all workloads. 
			On average, the error probability is increased by 32.5x.
			Regarding retention failure probability, thermal stability factor ($\Delta$) is the only parameter affected by PVs.
			Although PVs reduce the retention failure probability for some STT-MRAM cells and increase it for other cells, the total effect of PVs on retention failure probability of the cache result in a significant increase in its probability for all workloads.

	The effects of PVs on read disturbance probability is shown in Fig. \ref{fig:10}.
	On average, the error probability is increased by 9.0x when considering the PV effects.
	The results show that the effects of PVs on error probability are almost similar for all workloads.
	Read current ($I_{read}$), critical switching current ($I_{C_{0}}$), and thermal stability factor ($\Delta$) are under PVs and can oppositely affect the vulnerability of an STT-MRAM cell to read disturbance.
	Although the PVs can increase the error probability in some cells and decrease in other cells, the overall effect of PVs on STT-MRAM cache is to significantly increase the read disturbance error probability.

\begin{figure*}[t]\vspace{-10pt}
				\centering
				\includegraphics[width=1\linewidth]{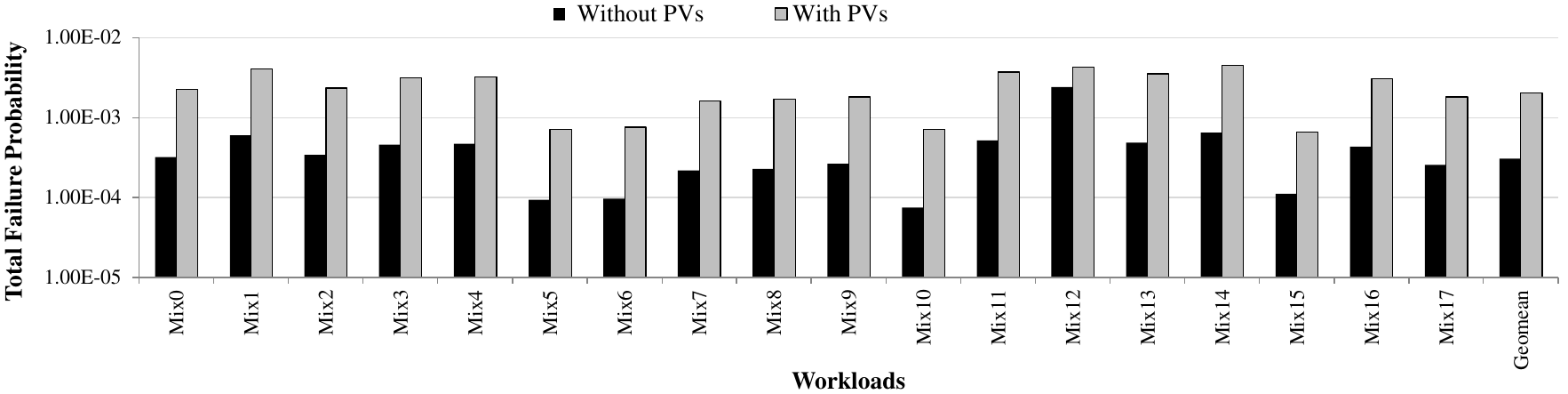}\vspace{-8pt}%
				\caption{Effect of workloads on total failure probability without/with considering PV effects.}\vspace{-10pt}
				\label{fig:12}
			\end{figure*}

			\begin{figure*}[b]\vspace{-10pt}
				\centering 	
				\subfloat[]{\includegraphics[width=1\linewidth]{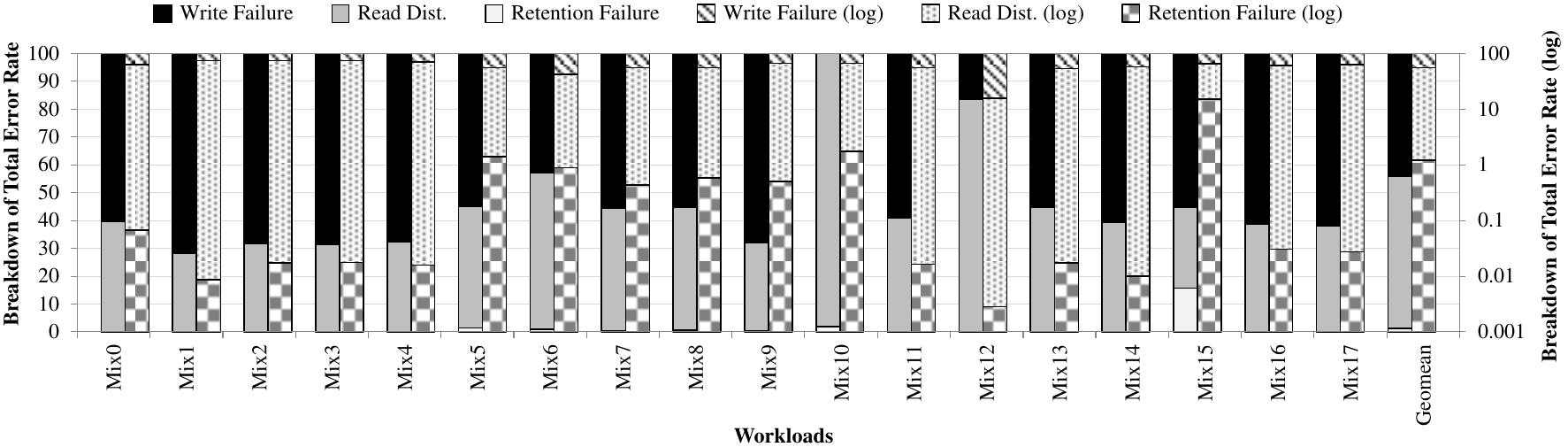}}\vspace{-8pt}
				\hfill
				\subfloat[]{\includegraphics[width=1\linewidth]{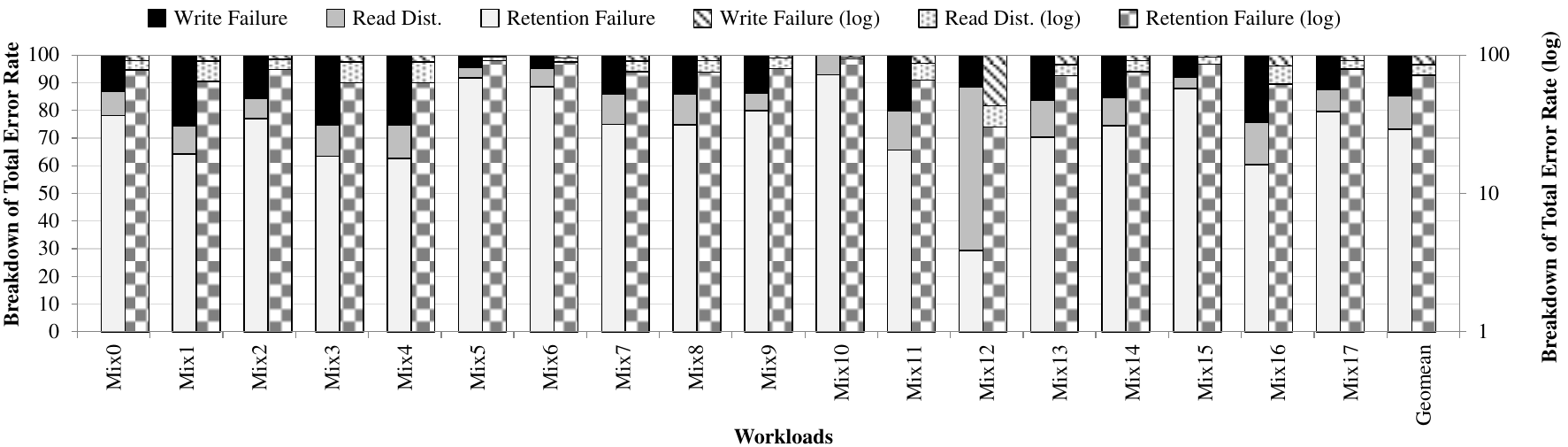}}\vspace{-3pt}
				\caption{Breakdown of total error rate to retention failure, read disturbance, and write failure (a) retention failure probability in only vulnerable intervals, (b) retention failure probability in all intervals.}
				\label{fig:13}
			\end{figure*}

			The probability of write failure affected by PVs in conjunction to that without PVs is depicted in Fig. \ref{fig:11}.
			The results show an average of 16.7x increase in write failure probability when considering the PV effects.
			The effect of PVs on a single STT-MRAM cell is either to increase or decrease the write failure probability depending on the deviations of thermal stability factor ($\Delta$), write current ($I_{write}$), and critical switching current ($I_{C_{0}}$) from their nominal values.
			However, the overall consequence of PVs on the write failure of a STT-MRAM cache is to significantly increase the probability of this error.
			
			\vspace{-8pt}
			\subsection{Total Error Rate}

			Fig. \ref{fig:12} depicts the total cache failure probability including retention failure, read disturbance, and write failure rates, without and with PVs consideration.
			On average, the total failure probability is 3.1$\times$$10^{\text{-4}}$ and is increased to 2.0$\times$$10^{\text{-3}}$ when considering PV effects.
			This observation implies more than 6.5x increase in total failure probability by the PVs.
			Without considering PVs, the minimum and maximum failure probabilities are observed in $Mix10$ and $Mix12$, respectively, which indicates a large gap between the corner cases.
			The maximum failure probability observed in $Mix12$ workload is by 32.0x higher than the minimum failure probability observed in $Mix10$ workload.
			The variation between the failure probabilities of workloads when considering PV effects is much lower than that of not including PVs.
			The maximum failure probability, which is observed in $Mix14$ workload (4.5$\times$$10^{\text{-3}}$) is by 6.8x higher than the minimum failure probability, which is observed in $Mix15$ workload (6.6$\times$$10^{\text{-4}}$).

			The source of large values in total cache failure probability is because of high write failure and read disturbance probabilities, as observed in Fig. \ref{fig:10} and Fig. \ref{fig:11}.
			The read disturbance and write failure probabilities of the cache are affected by the circuit- and physical-level parameters that determine the error rate of a single STT-MRAM cell.
			Error rate is exponentially affected by changing these parameters, e.g., write current and pulse width, read current and pulse width, and $\Delta$, as previously observed in (\ref{eq:5}) and (\ref{eq:6}).
			The goal of this study is \textit{not} to focus on the absolute error rate of STT-MRAM caches, which is strongly technology-dependent and is largely affected by design- and manufacturing-time parameters adjustment.
			Our goal, regardless of the per-cell error rate of the STT-MRAM cache, which is fixed and measurable through manufacturing process, is to demonstrate that: \textbf{a)} total cache error rate is largely affected by workloads behavior and a significant variation exists in different workloads; \textbf{b)}~\textit{none} of the three sources of errors can be ignored, since they have considerable contribution in total cache error rate; \textbf{c)} the workload behavior causes a large variation in the contribution of three sources of errors in total error rate.

			Fig. \ref{fig:13} shows the breakdown of total failure probability for all workloads in two scenarios for retention failure.
			As mentioned, retention failure occurred in an interval between a read/write access followed by a write access is overwritten by an updated data and can be ignored.
			Fig. \ref{fig:13}(a) and Fig. \ref{fig:13}(b) depict the breakdown of total failure probability when excluding the mentioned intervals in retention failure and when including them, respectively.
			According to Fig. \ref{fig:13}(a), read disturbance and write failure contribute in the majority of total failure probability, whereas the average contribution of retention failure is only 1.2\%.
			Due to negligible contribution of retention failure in this scenario, we depict the results in both linear and logarithmic scales.
			On average, 54.6\% of total failure probability is due to read disturbance while write failure causes 44.2\% of errors.
			A maximum of 71.6\% contribution for write failure is observed in $Mix1$ workload, while the minimum value is less than 0.05\% in $Mix10$ workload.
			The contribution of read disturbance varies from 28.4\% in $Mix1$ workload to 98.2\% in $Mix10$ workload.
			A significantly larger variation is observed in the contribution of retention failure, in which the minimum and maximum values are 0.003\% and 15.7\% in $Mix12$ and $Mix15$ workloads, respectively.
			It is noteworthy that the results are for an L2-cache and the contribution of retention failure will significantly increase in lower-level caches, e.g., L3-caches, due to lower frequency of read/write accesses and larger cache size.

			 Fig. \ref{fig:13}(b) depicts the breakdown of total failure probability considering all idle intervals of cache blocks in retention failure probability calculation.
			 The contribution of retention failure is as high as 73.1\%, on average, while read disturbance and write failure contribute for 12.2\% and 14.7\% of total errors, respectively.
			 The significant increase in the contribution of retention failure compared with that in Fig. \ref{fig:13}(a) implies that the cache blocks spend a large fraction of their idle time in intervals that are waiting to be written.
			 These intervals are the times between a read followed by a writeback access or eviction of the blocks.
			 Read disturbance contributes from 3.7\% in $Mix5$ workload to as high as 59.0\% in $Mix12$ workload.
			 The minimum and maximum value for write failure contribution are 0.002\% and 25.6\% in $Mix10$ and $Mix1$ workloads, respectively.

			\section{Discussion and Guidelines} 
			Several techniques have been presented in the literature to overcome each source of errors in STT-MRAM memories.
			Increasing write current and write pulse width, decreasing the thermal stability factor, reading and verifying data after each write (write-read-verify), and employing ${Error}$-${Correcting~Codes}$ (ECCs) are the technique to overcome the write failures~\cite{nowak2016dependence,sun-TMAG-12,sun2011design,lakys2012self,Pajouhi2016JETC,naeimi2013intel,ZAZADTE,ZAZADTPDS,ahn2013selectively}.
			Reducing read current and read pulse width, increasing thermal stability factor, overwriting after each read, and employing ECCs are among the techniques that tackle read disturbance~\cite{wang2015selective,Ran2016JSA,Na-TCAS-II-16, maddah2015cafo,14-zazad-eken2014novel}.
			Vulnerability to retention failure is reduced by increasing thermal stability factor, employing ECCs, and memory scrubbing\cite{naeimi2013intel,smullen2011relaxing,mittal2017survey}.

			There are three challenges in designing a reliable STT-MRAM cache based on the existing techniques.
			First, these techniques overcome the errors independently and three error mitigation techniques need to be employed for three sources of errors.
			Without including the workload-dependency of each error type and its contribution in total error rate, each technique should be configured for the worst case error rate.
			This pessimistic assumption leads to an overdesigned cache with significant protection overheads.
			For example, several studies presented ECCs for each error type and to overcome all three sources of errors, three kinds of coding need to be employed for each cache block.
			
			Second, techniques for protecting against one error may increase the rate of other error type(s).
			Reduction in $\Delta$ for decreasing write failure increases the rate of read disturbance and retention failure.
			Increasing $I_{write}$ and $t_{write}$ worsens the probability of MTJ barrier breakdown in addition to imposing significant energy consumption and delay overheads.
			Reducing $I_{read}$ and $t_{read}$ increases the sensing error probability.
			In addition, memory scrubbing increases the rate of read disturbance.
			
			Third, some of these techniques are in contradiction to each other and cannot be employed jointly.
			Increasing $\Delta$ of STT-MRAM cells for reducing read disturbance and retention failure rate is in contrast with the technique of decreasing $\Delta$ for reducing write failure rate.
			Overwriting a data block after each read for correcting read disturbance errors cannot be integrated with write-read-verify technique.

			Our formulations and framework help to estimate the cache reliability before and after employing error mitigation techniques and to determine the level of protection required for each error type based on its effects on the total error rate.
			Based on our observations, there is a large difference between the contribution of error types in total cache error rate and even in the rate of each error in various scenarios.
			These observations provide a roadmap to optimize the given budget of the cache protection for maximizing the reliability or to achieve the required level of reliability with the minimum cost.
			They can also help to design cache configurations that adaptively utilize error mitigation techniques at runtime based on the current behavior of different sources of errors.

			\section{Conclusions} 

			Unreliability of emerging STT-MRAM caches due to retention failure, read disturbance, and write failure is a major design concern.
			Vulnerability of cache blocks to these sources of errors strongly depends on not only process variations, but also the cache access patterns and data content.
			On the other hand, the rate of different errors is differently and even oppositely affected by a workload behavior and process variations. 
			This paper formulated the vulnerability of STT-MRAM caches to all error types considering the conflicting effects of both PVs and workloads behavior on the rate of different errors.
			These formulations provide an analytical reliability exploration environment for STT-MRAM caches.
			In addition, we developed a system-level framework to empirically investigate the reliability of STT-MRAM caches for different device- to system-level configurations and various PVs and workload considerations.
			We integrated our framework with gem5 full-system simulator and investigated the effects of PVs and workloads on the cache reliability.
			The results for a shared L2-cache in a quad-core processor showed that PVs increase the cache error rate by 6.5x, on average, while this rate varies for an average of 32.0x in different workloads.
			In addition, a large diversity is observed in the contribution of each error type in total vulnerability of STT-MRAM cache.
			Our formulations and framework help system architects to optimize the error mitigation techniques based on each error behavior and to design cost-efficient highly reliable STT-MRAM caches.
\bibliographystyle{IEEEtran}
			\bibliography{IEEEabrv,references}
			%
			%
			%
			
			%
				\begin{IEEEbiography}[{\includegraphics[width=1in,height=1.25in,clip,keepaspectratio]{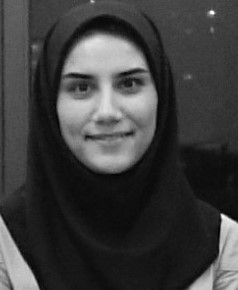}}]{Elham Cheshmikhani}
				received the B.Sc. degree in computer engineering from Iran University of Science and Technology (IUST) and the M.Sc. degree in computer engineering from Amirkabir University of Technology (Tehran Polytechnic-AUT), Tehran, Iran, in 2011 and 2013, respectively. She is currently a PhD candidate in computer engineering at Sharif University of Technology (SUT), Tehran, Iran.
She was a member of Design and Analysis of Dependable Systems (DADS) and Dependable Systems Laboratory (DSL) at AUT and SUT from 2011 to 2015 and 2015 to 2017, respectively. She is a member of Data Storage, Networks \& Processing Laboratory (DSN) since 2017 at SUT. Her research interests include emerging nonvolatile memory technologies, dependability analysis, fault tolerance, and storage systems.
		\end{IEEEbiography}
						
			\begin{IEEEbiography}[{\includegraphics[width=1in,height=1.25in,clip,keepaspectratio]{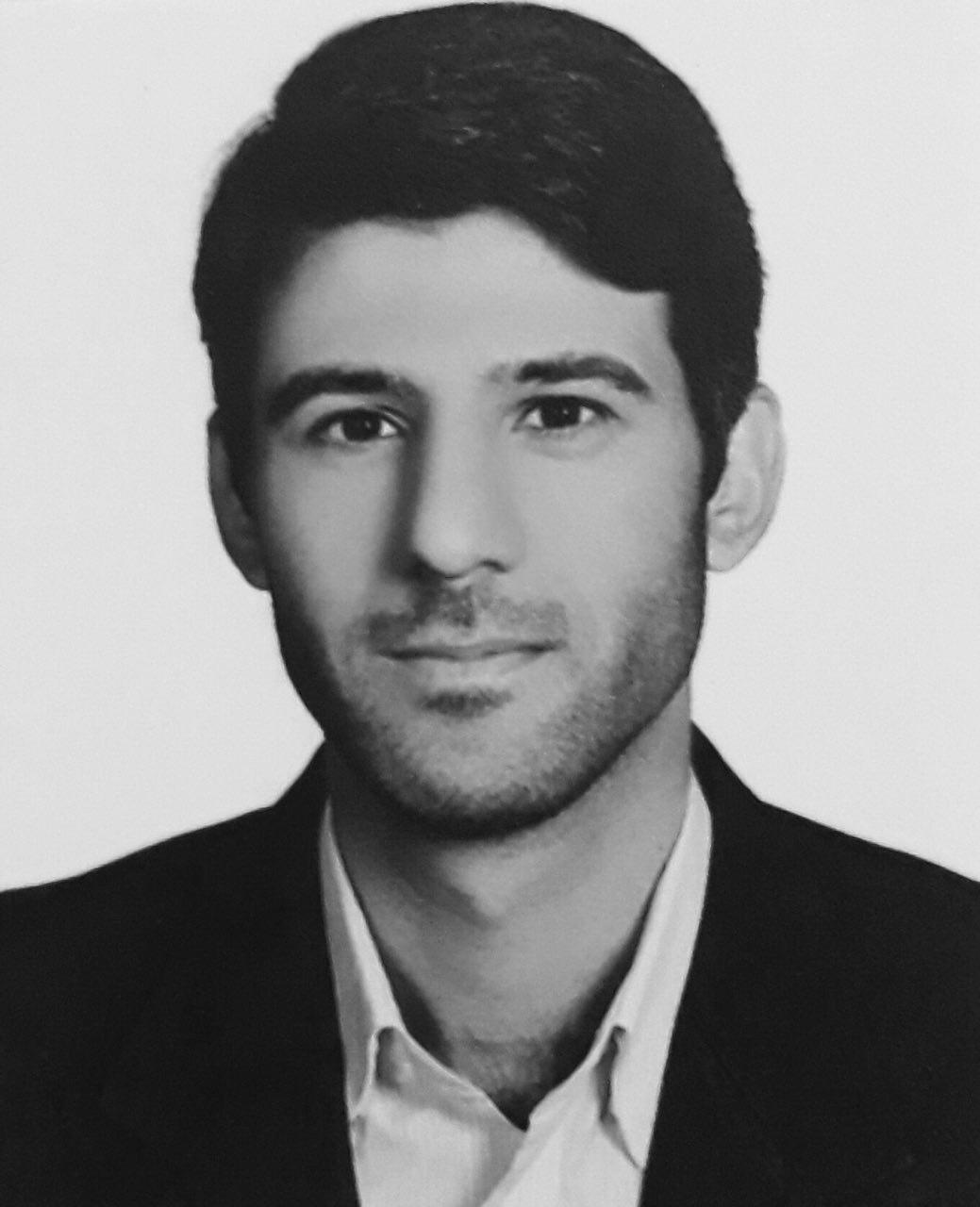}}]{Hamed Farbeh}
			(S'12) received the B.Sc., M.Sc., and PhD degrees in computer engineering from Sharif University of Technology (SUT), Tehran, Iran, in 2009, 2011, and 2017, respectively.
				He was a member of the Dependable Systems Laboratory (DSL) at SUT from 2007 to 2017 and the head of DSL from April 2017 to February 2018. 
				He is currently the faculty member of the Department of Computer Engineering and Information Technology, Amirkabir University of Technology (Tehran Polytechnic-AUT), Tehran, Iran. 
				He is also the head of Internet-of-Things (IoT) research center at AUT.
				He was with the Embedded Computing Laboratory (ECL), KAIST, Daejeon, South Korea, as a Visiting Researcher from October 2014 to May 2015 and collaborated with the Institute of Research for Fundamental Sciences (IPM), Tehran, Iran, as Postdoc fellow from May 2017 to January 2018.
				His current research interests include reliable memory hierarchy, reliability challenges in emerging memory technologies, cyber-physical systems. He was the IEEE student member from 2012 to 2017.
%
			\end{IEEEbiography}
%
%
%
%

			
			\begin{IEEEbiography}[{\includegraphics[width=1in,height=1.25in,clip,keepaspectratio]{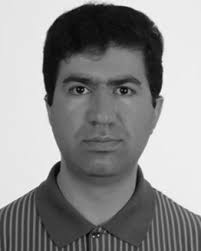}}]{Hossein Asadi}
				(M'08, SM'14) received the B.Sc. and M.Sc. degrees in computer engineering from the SUT, Tehran, Iran, in 2000 and 2002, respectively, and the Ph.D. degree in electrical and computer engineering from Northeastern University, Boston, MA, USA, in 2007. 
He was with EMC Corporation, Hopkinton, MA, USA, as a Research Scientist and Senior Hardware Engineer, from 2006 to 2009. From 2002 to 2003, he was a member of the Dependable Systems Laboratory, SUT, where he researched hardware verification techniques. From 2001 to 2002, he was a member of the Sharif Rescue Robots Group. He has been with the Department of Computer Engineering, SUT, since 2009, where he is currently a tenured Associate Professor. He is the Founder and Director of the \emph{Data Storage, Networks, and Processing} (DSN) Laboratory, Director of Sharif \emph{High-Performance Computing} (HPC) Center, and the President of Sharif ICT Innovation Center. He spent three months in the summer 2015 as a Visiting Professor at the School of Computer and Communication Sciences at the Ecole Poly-technique Federele de Lausanne (EPFL). He is also the co-founder of HPDS corp., designing and fabricating midrange and high-end data storage systems. He has authored and co-authored more than eighty technical papers in reputed journals and conference proceedings. His current research interests include data storage systems and networks, solid-state drives, operating system support for I/O and memory management, and reconfigurable and dependable computing.
Dr. Asadi was a recipient of the Technical Award for the Best Robot Design from the International RoboCup Rescue Competition, organized by AAAI and RoboCup, a recipient of Best Paper Award at the 15th CSI International Symposium on \emph{Computer Architecture \& Digital Systems} (CADS), the Distinguished Lecturer Award from SUT in 2010, the Distinguished Researcher Award and the Distinguished Research Institute Award from SUT in 2016, and the Distinguished Technology Award from SUT in 2017. He is also recipient of Extraordinary Ability in Science visa from US Citizenship and Immigration Services in 2008. He has also served as the publication chair of several national and international conferences including CNDS2013, AISP2013, and CSSE2013 during the past four years. Most recently, he has served as a Guest Editor of IEEE Transactions on Computers, an Associate Editor of Microelectronics Reliability, a Program Co-Chair of CADS2015, and the Program Chair of CSI National Computer Conference (CSICC2017). 
								
			\end{IEEEbiography}

			\end{document}